\documentclass[%
 reprint,
 amsmath,amssymb,
 aps,
]{revtex4-2}

\usepackage{graphicx}
\usepackage{dcolumn}
\usepackage{bm}
\usepackage{amsmath}
\usepackage{amssymb}
\usepackage{txfonts}
\usepackage{mathrsfs}
\usepackage{xcolor} 
\usepackage{bbold}

\usepackage[mathlines]{lineno}

\begin{document}

\title{Spin-orbit coupled mean-field Bose gas at finite temperature}
\author{Pawel Jakubczyk }
\affiliation{Institute  of Theoretical Physics, Faculty of Physics, University of Warsaw, Pasteura 5, 02-093 Warsaw, Poland}
\author{Marek Napiórkowski }
\affiliation{Institute  of Theoretical Physics, Faculty of Physics, University of Warsaw, Pasteura 5, 02-093 Warsaw, Poland}

\date{\today}

\begin{abstract} 
We consider the spin-orbit coupled Bose gas with repulsive mean-field   interparticle interactions. We analyze the phase diagram of the system varying the temperature $T>0$, the chemical potentials, as well as interparticle and spin-orbit interaction couplings. Our results indicate that, for Rashba- and Weyl-type spin-orbit couplings, condensates featuring ordering wavevector $\vec{Q}\neq \vec{0}$ are fragile with respect to thermal fluctuations and, at $T>0$, the only stable thermodynamic phases involving the Bose-Einstein condensate (BEC) are those of uniform type with $\vec{Q}=\vec{0}$. On the other hand, presence of the spin-orbit coupling  stabilizes the $\vec{Q}=\vec{0}$ BEC state at any dimensionality $d>1$ and modifies either the order or the universality class of the corresponding phase transition. We emphasize the singular nature of the limit of vanishing spin-orbit interaction coupling $v$, sizable shifts of the phase boundaries upon varying $v$, as well as the role of the relative magnitudes of the interparticle interaction  couplings for the character of the condensation transition.

\end{abstract}

\maketitle


\section{Introduction}
Recent years witnessed substantial interest in Bose systems,  where internal (pseudo-spin) degrees of freedom of neutral particles become coupled to their momenta.  Such ''synthetic'' spin-orbit (S-O) coupled setups can be realized by exposing the Bose gas to spatially nonuniform laser fields and lead to degenerate ground states characterized by non-zero ordering wavevectors,  sometimes referred to as supersolid states \cite{Boninsegni_579}. Numerous studies addressed the zero-temperature ($T$=0) phase diagrams of such spin-orbit coupled Bose-Einstein condensates as well as their excitation spectra, for reviews see \cite{Dalibard_2011, Galitski_2013, Zhou_2013, Goldman_2014, Zhai_2015, Zhang_2016,  Recati_2022}. Attention has also been paid to their finite temperature properties, in particular the stability of the modulated states with respect to thermal fluctuations \cite{Stanescu_2008, Barnett_2012, Barnett_2012_erratum, Ozawa_2012_2, Cui_2013, Yu_2013, He_2013, Liao_2014, Roy_2014, Hickey_2014, Mardonov_2015, Liao_2015, Galteland_2016, Wu_2017, Kawasaki_2017, Yang_2021} depending on the system dimensionality, presence or absence of interparticle interactions, and the particular form of the S-O coupling. This question is well-grounded considering that breaking translational invariance leads to enhanced Goldstone fluctuations, which may well destabilize such long-range ordered states. We in particular point out that in the related context of neutral fermionic superfluids featuring the ordering wavevector $\vec{Q}\neq \vec{0}$ (known as FFLO states) thermal fluctuations are believed to give rise to  instabilities \cite{Shimahara_1998, Radzihovshy_2009,  Radzihovsky_2011, Yin_2014,  Jakubczyk_2017, Wang_2018, Zdybel_2021, Pini_2021,Pini_2023} of the long-range ordered modulated phases both in dimensionality $d=2$ and $d=3$. 
Interestingly, the situation seems a bit less clear for Bose systems, where one encouters different literature claims - for example compare Refs.~\cite{Barnett_2012, Barnett_2012_erratum} and Ref.~\cite{Cui_2013}. Another general interesting question concerns the possibility of realizing condensation in states which do not correspond to the global minimum of the spectrum of the non-interacting component of the Hamiltonian. Such possibility arises because the fluctuation spectrum above a putative ground state \emph{does} depend on the state in question and states of lower energy may turn out less stable when fluctuations are considered on top of them.  

In this work we address the stability of the different thermodynamic phases involving the BEC as well as the structure of the phase diagram and the order of the phase transitons, employing an exactly solvable model of bosons, which combines particular realizations of the S-O coupling and  interparticle interactions included at the mean-field level. Our results show that for Rashba-type S-O coupling the long-range ordered modulated BEC (characterized by $\vec{Q}\neq\vec{0}$) does not represent a stable thermodynamic phase in dimensionality $d=3$. Similar conclusion holds for the Weyl-type S-O coupling. However (for the Rashba case) the S-O interaction stabilizes the  uniform long-range ordered BEC state ($\vec{Q}=0$), such that it may exist at $T>0$ also in dimensionality $d=2$.  Concerning the  condensation transition between the normal and uniform BEC states, our results indicate that, for certain parameter ranges, introducing the S-O coupling changes the order of the transition by driving it first-order. On the other hand, in cases, where the transition remains second-order, its universality class becomes modified and the critical temperature is elevated as compared to the case of vanishing S-O coupling.   

The present paper is organized as follows: In Sec.~II we introduce the proposed model and discuss its exact solution. In particular we identify the BEC phases which may, and which may not, be realized in dimensionalities $d\in\{2,3\}$. We also discuss an exact analytical relation  between the density and temperature at condensation in situations, where the transition is continuous. We compare our results to the case of vanishing S-O coupling. In Sec.~III we numerically construct the phase diagrams and assess the impact of the S-O coupling on the order of the phase transitions and the phase boundary shifts. We identify qualitatively different situations depending on the relative magnitue of the inter- and intra-species interaction couplings. 
We summarize the paper and give a perspective for future studies in Sec.~IV.

\section{Model and its solution}
We consider the gas of  Rashba S-O coupled bosons described by the following Hamiltonian:
\begin{align}    \hat{\mathcal{H}}=\sum_{\vec{p},\alpha,\beta}\langle\vec{p},\alpha|\hat{\mathcal{H}}^{(1)}|\vec{p},\beta\rangle \hat{a}^\dagger_{\vec{p},\alpha} \hat{a}_{\vec{p},\beta} +\hat{\mathcal{H}}_{int}^{MF}\;. 
\label{Hamiltonian}
\end{align}
Here $\{\hat{a}^\dagger_{\vec{p},\alpha}\}, \{\hat{a}_{\vec{p},\beta}\}$ are bosonic creation/annihilation operators, and  
\begin{align}
\label{R01}
    \hat{\mathcal{H}}^{(1)}=\frac{\vec{p}^2}{2m}\hat{\mathbb{1}}_\sigma-v\left(p_1\hat{\sigma}_2+p_2\hat{\sigma}_3\right)\;.
\end{align}
 The single-particle states are labeled by momentum $\vec{p}$ and the pseudospin 1/2 index ($\alpha$, $\beta\in\{\uparrow, \downarrow\}$),   $\hat{\sigma}_i$ are the Pauli matrices and $v\geq 0$ represents the spin-orbit coupling. The free part of the Hamiltonian of Eq.~(\ref{Hamiltonian}) is the same as the one analyzed e.g. in Ref.~\cite{Stanescu_2008}.
 For discussions of experimental realizations leading to this type of artificial pseudospin 1/2 bosons and the S-O coupling see e.g. Refs \cite{Osterloh_2005, Zhu_2006, Satija_2006, Stanescu_2007}. We do not specify the dimensionality $d$ of the system, such that $\vec{p}=(p_1,\,\dots, p_d)$, but restrict our analysis to $d\geq 2$. The system is contained in a hypercubic vessel of volume $V$ and we assume periodic boundary conditions, leading to translational invariance. For now we concentrate on the Rashba type S-O coupling, we will discuss the Weyl case later on (see Sec.~IIC). 
 The one-body component of the Hamiltonian of Eq.~(\ref{Hamiltonian}) can be diagonalized by a unitary transformation \cite{Stanescu_2008} $\hat{U}_{\vec{p};\sigma,\beta}$ (diagonal in momentum space), leading to   
 \begin{align}    
 \hat{\mathcal{H}}=\sum_{\vec{p},\sigma}E_{\vec{p},\sigma} \hat{A}^\dagger_{\vec{p},\sigma} \hat{A}_{\vec{p},\sigma} +\hat{\mathcal{H}}_{int}^{MF}\;,  
\label{Hamiltonian_d}
\end{align}
 where $\hat{A}_{\vec{p},\sigma} =\sum_\beta\hat{U}_{\vec{p};\sigma,\beta}\, \hat{a}_{\vec{p},\beta}$. 
 The single-particle spectrum reads: 
\begin{align}
\label{dispersion}
E_{\vec{p},\sigma}=\frac{\vec{p}^2}{2m}+\sigma v\sqrt{p_1^2+p_2^2}\;,  
\end{align} 
where $\sigma\in\{\pm 1\}$. For $v>0$ the upper branch  ($\sigma=+1$) exhibits a cuspy minimum located at $\vec{p}=\vec{0}$ with $E_{\vec{0},+}=0$. The lower branch ($\sigma=-1$)
has a degenerate minimum $E_{\vec{p}_0,-}=-mv^2/2$ occurring for $\sqrt{p_1^2+p_2^2}=mv=:p_0$ and $p_i=0$ for $i>2$. 

In the presently considered model, the term $\hat{\mathcal{H}}_{int}^{MF}$ in Eq.~(\ref{Hamiltonian_d}) and Eq.~(\ref{Hamiltonian}) accounts for interactions between the quasiparticles represented by the operators $\{\hat{A}_{\vec{p},\sigma}^\dagger\},\{\hat{A}_{\vec{p},\sigma}\}$. It is assumed to take the following mean-field form: 
\begin{align}
    \hat{\mathcal{H}}_{int}^{MF}= \sum_{\sigma,\sigma'}\frac{a_{\sigma \sigma'}}{2V}\hat{N}_\sigma \hat{N}_{\sigma'}\;, 
    \label{Hint}
\end{align}
where the interaction couplings $a_{\sigma \sigma'}$ are positive and $\hat{N}_\sigma= \sum_{\vec{p}}\hat{A}^\dagger_{\vec{p},\sigma} \hat{A}_{\vec{p},\sigma}$ is the operator representing the total number of quasiparticles occupying the $\sigma$ band.

In what follows, we shall denote the interaction couplings $a_{\sigma, \sigma}$ as $a_\sigma$, and $a_{+,-}=a_{-,+}$ as $a_{\pm}$.  The mean-field interaction specified in Eq.~(\ref{Hint}) can be derived from a realistic repulsive two-body potential acting between the  (quasi)particles in the so-called Kac limit, where the interaction becomes increasingly weak and long-ranged (see e.g.\cite{Hemmer_1976}). For the one-component case without the S-O coupling this model is recognized as the imperfect Bose gas and was first discussed in Ref. \cite{Davies_1972} (see also Refs.~\cite{Buffet_1983, Berg_1984, Lewis_1985, Smedt_1986, Zagrebnov_2001}). Its phase diagram as well as critical properties were fully clarified in Refs.~\cite{Napiorkowski_2011, Napiorkowski_2013, Diehl_2017,  Jakubczyk_2018}. Concerning Bose-Einstein condensation, the model belongs to the universality class of the spherical (Berlin-Kac) model \cite{Berlin_1952}, corresponding also to the limit $N\to \infty$ of $O(N)$-symmetric models \cite{Stanley_1968, Moshe_2003}. The case of two-component imperfect Bose mixture was addressed only recently \cite{Jakubczyk_2024} yielding a relatively complex phase diagram featuring in particular both first- and second-order condensation transitions and rich multicritical behavior. 

In regard to the present physical situation, we observe that the system defined in Eqs.~(\ref{Hamiltonian_d}, \ref{dispersion}, \ref{Hint}) may be understood as a two-component mixture of quasiparticles characterized by the dispersions given by Eq.~(\ref{dispersion}) and interacting via the mean-field repulsion introduced by Eq.~(\ref{Hint}). As such, it exhibits a degree of affinity to the system studied in Ref. \cite{Jakubczyk_2024} and one may efficiently adapt the formalism developed therein to treat the present setup. 
However, as we shall discuss in detail, the limit $v\to 0$ is singular and the properties of the system at $v>0$ and $v=0$ are entirely different. Some technical differences between the  $v>0$ and $v=0$ cases are discussed in Sec.~III.

We begin the analysis by noting that the grand-canonical partition function of the present model can be obtained along the line developed for the standard Bose mixture \cite{Jakubczyk_2024} and reads: 
\begin{align}
    \Xi(T,V,\mu_+,\mu_-)=c\frac{\beta V}{\sqrt{|D|}}\int_{\Gamma_1} dt_1 \int_{\Gamma_2} dt_2e^{-V\Phi(t_1,t_2)}\;,  
    \label{Xigen}
\end{align}
where $D=a_+ a_- -a_{\pm}^2$ is the interaction matrix determinant, $\beta=(k_BT)^{-1}$, $c$ is a numerical constant depending on the sign of $D$, and the complex integration  contours $\Gamma_1$ and $\Gamma_2$ are parallel or perpendicular to the imaginary axes, depending on the sign of $D$. The presence of the volume factor in the argument of the exponential in the integrand in Eq.~(\ref{Xigen}) assures that the saddle-point evaluation of the integral becomes exact for  $V\to\infty$. In consequence, the mean-field treatment  of the model represents the exact solution in the thermodynamic limit. This feature of the analysis follows from the assumed  extremely weak and long-ranged interparticle interactions described by Eq.~(\ref{Hint}), which is specific to mean-field models. On the other hand we note \cite{Jakubczyk_2024}, that (at least for $v=0$) the predictions delivered by the present model are closely related (and in some aspects equivalent) to those of the Hartree-Fock treatment of the dilute Bose gases with short-range forces.

The function $\Phi(t_1,t_2)$ in Eq.~(\ref{Xigen}) is given as: 
\begin{align} 
\label{Phi}
    \Phi(t_1,t_2)=& -\frac{\beta}{2a_+}(t_1-\mu_+)^2-\frac{\beta}{2a_-'}(t_2-\mu_-')^2 \\&- \frac{1}{V}\log\Xi_0^{(+)}(T,V,t_1) - \frac{1}{V}\log\Xi_0^{(-)}(T,V,\frac{a_\pm}{a_+}t_1+t_2)\;,\nonumber
\end{align}
where $\Xi_0^{(\sigma)}(T,V,\mu_\sigma)$ represents the grand canonical partition function of a system of non-interacting quasiparticles characterized by the dispersion $E_{\vec{p},\sigma}$ given by Eq.~(\ref{dispersion}). To shorten notation we introduced $\mu_-'=\mu_--a_{\pm}\mu_+/a_+$ and $a_-'=a_--a_{\pm}^2/a_+$, where $\mu_{+}$ and $\mu_{-}$ denote the corresponding  chemical potentials. We emphasize that, on the technical level, the key difference between the present case and the standard Bose mixture in absence of the S-O coupling is encoded in the dispersion relation. In particular for $v=0$ we recover the system considered in Ref.~\cite{Jakubczyk_2024}. For the detailed derivation of Eq.~(\ref{Xigen},\ref{Phi}) we refer to \cite{Jakubczyk_2024}.

In view of the structure of Eq.~(\ref{Xigen}) the evaluation of the partition function $\Xi(T,V,\mu_+,\mu_-)$ in the thermodynamic limit amounts to finding $\Phi(\bar{t}_1, \bar{t}_2)$, where $(\bar{t_1},\bar{t_2})$ represents the dominant saddle point of $\Phi(t_1,t_2)$. In particular, the grand canonical free energy density 
\begin{align}
    \omega=-\beta^{-1}V^{-1}\log\Xi\longrightarrow \beta^{-1}\Phi (\bar{t_1},\bar{t_2})
\end{align}
in the thermodynamic limit $V\to\infty$. 
Considering that  
\begin{align} 
    \Xi_0^{(\sigma)}(T,V,\mu_{\sigma}) = \prod_{\vec{p}}\frac{1}{1-e^{-\beta(E_{\vec{p},\sigma}-\mu_\sigma)}}\;, 
\end{align}
we may (for $V\to\infty$) transform $\Xi_0^{(\sigma)}$ to the following forms: 
\begin{align} 
\label{Xi0minus}
    \frac{1}{V}\log\Xi_0^{(-)}=&\frac{1}{\lambda^d}\left[ g_{\frac{d+2}{2}}(z_-)+ \tilde{v}\left(\sqrt{\pi}\,g_{\frac{d+1}{2}}(z_-) + f_1^{(d)}(\tilde{v},z_-)\right) \right] \\ -&\kappa\log(1-z_-)\;,\nonumber
\end{align} 
and 
\begin{align} 
\label{Xi0plus}
    \frac{1}{V}\log\Xi_0^{(+)}= \frac{1}{\lambda^d}\left[g_{\frac{d+2}{2}}(z_+)- f_2^{(d)}(\tilde{v},z_+) \right]-\log(1-z_+)\;.
\end{align} 
Above we defined
\begin{align}
    f_1^{(d)}(\tilde{v},z):=\int_0^{\tilde{v}^2}dx\left(-\tilde{v}^{-1}+x^{-1/2}\right)g_{\frac{d}{2}}(z e^{-x})\;, 
    \label{f1def}
\end{align} 
and 
\begin{align}
    f_2^{(d)}(\tilde{v},z):=\int_0^{\infty}dx\left(1+\frac{x}{\tilde{v}^2}\right)^{-1/2}g_{\frac{d}{2}}(z e^{-x})\;. 
    \label{f2def}
\end{align} 
We also introduced 
\begin{align}
\label{param01}
    \tilde{v}:=\sqrt{\frac{mv^2}{2k_BT}}\;,\;\;\; z_-:=e^{\beta \mu_- + \,\tilde{v}^2}\;,\;\;\; z_+:=e^{\beta\mu_+}\;.  
\end{align}
The quantity $\kappa$ in Eq.~\eqref{Xi0minus} is 
the surface area of the $(d-1)$-dimensional unit sphere 
and the thermal de Broglie length $\lambda$ and the Bose functions $g_\alpha(z)$ are as usual defined by 
\begin{align}
    \lambda=\frac{h}{\sqrt{2\pi m k_B T}}\;\;\;\; \textrm{and} \;\;\;\; g_\alpha(z)=\sum_{k=1}^{\infty}\frac{z^k}{k^\alpha} \,,
\end{align}
respectively. The limit of vanishing S-O coupling may at any time be recovered by taking $\tilde{v}\to 0$. 

We proceed by inserting the expressions given by Eq.~(\ref{Xi0minus}) and (\ref{Xi0plus}) into Eq.~(\ref{Phi}) and evaluating the saddle-point equations: $\left.\partial\Phi/\partial t_1 \right|_{\{t_i=\bar{t}_i\}} = \left.\partial\Phi/\partial t_2 \right|_{\{t_i=\bar{t}_i\}}=0$ from which one obtains expressions for $\bar{t}_{i}(\mu_{+},\mu_{-},T,\nu)$, $i=1,2$.  The average densities $n_\sigma$ are then obtained from 
\begin{equation}
    n_\sigma = -\frac{\partial\omega}{\partial \mu_\sigma}=-\beta^{-1}\frac{\partial\Phi(\bar{t}_1,\bar{t}_2)}{\partial\mu_\sigma}\;,
\end{equation}
which leads to the following simple relation between $\bar{t}_i$ and $n_\sigma$:  
\begin{align} 
    n_+ =& -\frac{1}{a_+}(\bar{t_1}-\mu_+)+\frac{1}{a_-'}(\bar{t_2}-\mu_-')\frac{a_{\pm}}{a_+} \,,\nonumber \\
    n_- =& -\frac{1}{a_-'}(\bar{t_2}-\mu_-')  \;. 
    \label{tton}
\end{align} 
Using the above transformation one may eliminate $\bar{t}_1$ and $\bar{t}_2$ from the saddle-point equations in favor of $n_+$ and $n_-$. This way we obtain: 
\begin{align} 
\label{Densitieseq}
    n_+ \lambda^d &=g_{\frac{d}{2}}\left(e^{s_+}\right)- f_2^{(d-2)}\left(\tilde{v},e^{s_+}\right)+\frac{1}{V}\frac{e^{s_+}}{1-e^{s_+}} \,,\nonumber\\  
    n_- \lambda^d &=g_{\frac{d}{2}}\left(e^{s_-}\right)+ \sqrt{\pi}\,\tilde{v}\, g_{\frac{d-1}{2}} \left(e^{s-}\right)+\tilde{v}f_1^{(d-2)}(\tilde{v},e^{s_-}) \\ 
    &+\frac{\kappa}{V}\frac{e^{s_-}}{1-e^{s_-}}\;.     \nonumber
\end{align} 
The above equations are coupled via the quantities $s_+$ and $s_-$ defined as 
\begin{align}
\label{s01}
    s_+=\beta(\mu_+ -a_+ n_+ - a_{\pm}n_-)
    \end{align}
\begin{align}
\label{s02}
    s_-=\beta(\mu_- -a_- n_- - a_{\pm}n_+)+\tilde{v}^2\;
\end{align}
and their solution determines the average  densities of the quasiparticles. Note that $s_\pm<0$ and the terms $\sim 1/V$ in Eq.~(\ref{Densitieseq}) represent the condensate densities. These may give finite contributions only for $s_{\pm}$ approaching zero as $V\to \infty$. 
\subsection{Existence of BEC phases}
In this section we examine the analytical structure of Eqs ~(\ref{Densitieseq}) with the aim of identifying the admissible BEC phases depending on dimensionality $d$.  
We begin with the special case $\tilde{v}=0$, where all terms involving the $f_{i}^{(d-2)}$-functions vanish \cite{Jakubczyk_2024}.  We then recover the standard fact that condensation occurs for sufficiently low temperatures or high densities provided $d>2$. This follows from the property, that for $\alpha>1$ the function $g_\alpha(z)$ is bounded from above by its value at $z=1$ [$g_\alpha(1)=\zeta(\alpha)$, $\zeta$ denoting the Riemann zeta function], and is divergent for $z\to 1^-$ for $\alpha\leq 1$. Upon reducing $T$ (or increasing densities) the left-hand sides of Eqs ~(\ref{Densitieseq}) can be increased without bound. For $d\leq 2$, this unbounded growth of the left-hand side can be followed by increasing the value of $s_{\pm}$ in the argument of the Bose function $g_{\frac{d}{2}}$ the on the right-hand side of Eq.~(\ref{Densitieseq}) towards zero  such that Eqs ~(\ref{Densitieseq}) always remain fulfilled with $s_\pm<0$. On the other hand, for $d>2$ the function $g_{\frac{d}{2}}(z)$ is bounded and fulfillment of Eqs ~(\ref{Densitieseq}) at $T$ sufficiently low requires an additional contribution which is provided by the terms $\sim 1/V$. The  logic of this  argument is completely analogous as for the non-interacting Bose gas (see e.g. \cite{Ziff_1977, Kardar_book}). 

 The same reasoning may however be straightforwardly adapted to the case $\tilde{v}>0$. We consider the lower branch $E_{\vec{p},-}$ of the dispersion relation with minimum at $\sqrt{p_1^2+p_2^2}=m\nu$ first. By inspecting Eqs (\ref{Densitieseq}) we note the term involving $g_{\frac{d-1}{2}}(e^{s_-})$, which is unbounded for $s_-\to 0^-$ if $d\leq 3$. It can be easily checked that its singularity is not cancelled by the contribution involving $f_1^{(d-2)}$ in Eqs (\ref{Densitieseq}). This leads to the conclusion that the putative modulated condensed phase in the Bose gas involving Rashba S-O coupling may only be realized at $T>0$ for dimensionalities $d>3$. 

The situation is entirely different in regard to the particles occupying the upper branch $E_{\vec{p},+}$. Let us examine the function $f_2^{(d-2)}(\tilde{v},e^{s_+}) $ occurring in Eqs ~(\ref{Densitieseq}). We observe that for $\tilde{v}>0$ the dominant contribution to the integral defining $f_2^{(d-2)}(\tilde{v},e^{s_+})$ [see Eq.~(\ref{f2def})] comes from small values of the integration variable $x$. At $x$ large the integrand is exponentially suppressed due to $g_\alpha(z e^{-x})\sim z e^{-x}$ for $x$ large. We therefore expand the integrand in $x/\tilde{v}^2$ and integrate term by term. The first term of the expansion then cancels $g_{\frac{d}{2}}(e^{s_+})$ in Eqs ~(\ref{Densitieseq}) and we obtain 
\begin{equation} 
\label{n+asympt}
    n_+ \lambda^d \approx \frac{1}{2\tilde{v}^2}\, g_{\frac{d}{2}+1}(e^{s_+}) +\frac{1}{V}\frac{e^{s_+}}{1-e^{s_+}}\;.
\end{equation}
For $d > 0$ the function $g_{\frac{d}{2}+1}(e^{s_+})$ is bounded for $s_{+} \rightarrow 0^{-}$. In consequence, the presence of the S-O coupling allows for realizing uniform BEC both in $d=3$ and $d=2$. 

The above analysis points at  singular nature of the limit $\tilde{v}\to 0$. In $d=3$ shifting $\tilde{v}$ from zero to an arbitrarily small positive value entirely eliminates one of the BEC phases from the phase diagram. Contrary, in $d=2$, a BEC phase absent for $\tilde{v}= 0$ appears in the phase diagram for any $\tilde{v}>0$. The discontinuous nature of the phase diagram of the system at $v=0$ is of course a consequence of the fact, that the one-particle dispersion evolves non-analytically when $v$ is shifted from zero to a positive value. We also point out that, provided the BEC transition is continuous, the critical singularities differ for the two cases corresponding to $v=0$ and $v>0$. While for the former case, the critical exponents are known to represent the $O(N\to\infty)$ universality class,   in the latter case we expect them to be mean-field (Landau-like) for both $d=2$ and $d=3$. This follows from the structure of the saddle-point equations at condensation (for $|s_+|\ll 1)$ - compare in particular Eq.~(\ref{Densitieseq}) with $\tilde{v}=0$ and Eq.~(\ref{n+asympt}). Obviously, Eq.~(\ref{n+asympt}) appears analogous to Eq.~(\ref{Densitieseq}) with $\tilde{v}=0$ upon elevating dimensionality by two. The effective dimensionality $D=d+2$ is therefore larger (or equal) to the value 4 corresponding to the upper critical dimension. 

\subsection{Critical densities for condensation}
Under the assumption that densities evolve continuously upon varying the chemical potentials, it is straightforward to obtain a simple expression for the density $n_{+}^{c}(T)$ at which the BEC transition in the $(+)-$component takes place. Fulfillment of the above mentioned continuity requirement is not a priori guaranteed even for $\tilde{v}=0$ (see Ref.\cite{Jakubczyk_2024}), but is satisfied in some of the physically most interesting cases. We elucidate this point in Sec.~III. Adopting the continuity assumption for the time being, we observe that at the BEC transition we have $s_+=0$ and at the same time the condensate density is zero, such that the term $\sim 1/V$ in Eqs ~(\ref{Densitieseq}) vanishes. This yields 
\begin{equation}
    n_{+}^{(c)}(T)=\lambda^{-d}\left[\zeta\left(\frac{d}{2}\right)-\int_0^{\infty}dx\left(1+\frac{x}{\tilde{v}^2}\right)^{-1/2}g_{\frac{d}{2}-1}(e^{-x})\right] 
    \label{nc}
\end{equation} 
for $d>2$. We shall separately analyze the case $d=2$. The obtained expression has a very clear scaling structure. The entire dependence of  $n_+^{(c)}(T)\lambda^{d}$ on the system parameters occurs via the dimensionless quantity $\tilde{v}^2$, see Eq.~\eqref{param01}.
For $\tilde{v}\ll 1$ the first term dominates and one recovers the standard result $n_+^{(c)}(T)\approx \zeta (d/2) \lambda^{-d}\sim T^{d/2}$. In contrast, for $\tilde{v}\gg 1$ we expand the integrand in $x/\tilde{v}^2$ truncating at next to leading term, which yields 
$n_+^{(c)}(T)\approx \zeta(1+d/2)\lambda^{-d}/(2\tilde{v}^2)\sim T^{1+d/2}/v^2$. These two scaling regimes and the associated crossover is demonstrated for $d=3$ in Fig.~1.    
\begin{figure}
\includegraphics[width=8.5cm]{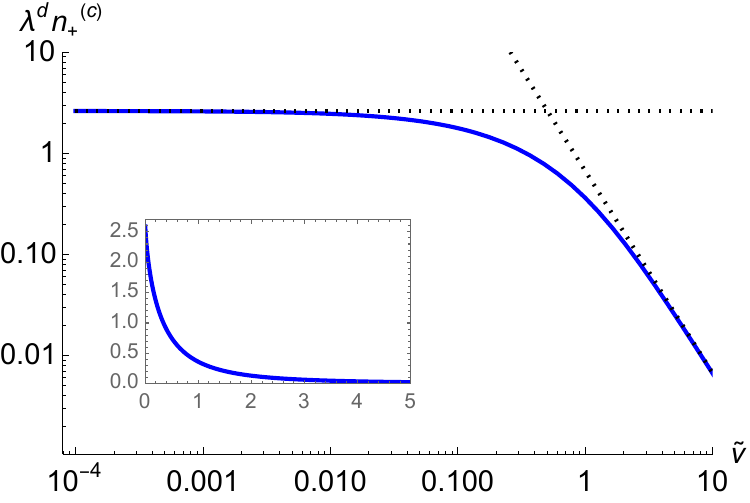}
\caption{The dimensionless critical density $n_+^{(c)}(T)\lambda^{d}$ as given by Eq.~(\ref{nc}) plotted as a function of $\tilde{v}$ in $d=3$. The two scaling regimes correspond to $n_+^{(c)}(T)\sim T^{3/2}$ (for $\tilde{v}\ll 1$) and $n_+^{(c)}(T)\sim T^{5/2}$ (for $\tilde{v}\gg 1$). The inset shows the same data plotted in the linear scale.}
\end{figure}   
The structure of Eq.~(\ref{nc}) immediately implies that presence of the S-O coupling always leads to  reduction of the critical density as compared to the case $\tilde{v}=0$. 

The case $d=2$ must be treated separately due to divergence of $\zeta(d/2)$ for $d \rightarrow 2$. To recover $n_+^{(c)}(T)$ in this situation, we consider Eqs~(\ref{Densitieseq}) neglecting the last terms, and evaluate numerically the limit $s_+\to 0^-$ for different values of $\tilde{v}$. The result is summarized in Fig.~2. 
\begin{figure}
\includegraphics[width=8.5cm]{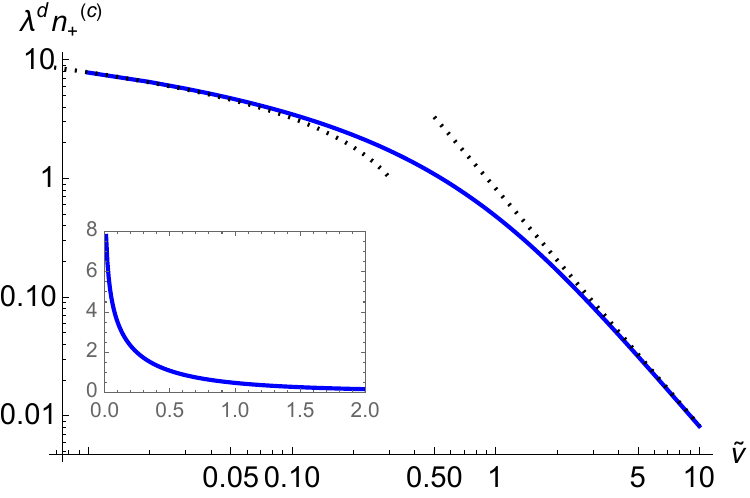}
\caption{The dimensionless critical density $n_+^{(c)}(T)\lambda^{d}$ plotted as a function of $\tilde{v}$ in $d=2$. The divergence for $\tilde{v}\to 0^+$ signals expelling the BEC phase out of the phase diagram in absence of the S-O coupling. The asymptotic formulae for $n_{+}^{(c)}(T)\lambda^{d}$ are given in Eq.~(\ref{asymptd2}). The inset shows the same data plotted in the linear scale.} 
\end{figure}     
As expected, $n_+^{(c)}$ remains finite for all values of $\tilde{v}>0$, which confirms the effect of stabilizing the condensate at $d=2$ by the S-O interaction, but diverges for $\tilde{v} \rightarrow 0$, as the condensate becomes depleted in absence of the S-O coupling. The analytic expressions describing the asymptotic behavior of $n_{+}(T)\lambda^2$ shown on Fig.~2 can be obtained from Eq.~\eqref{nc} by integrating by parts and subsequently substituting $d=2$ and  using $g_1(z)=-\log (1-z)$. We find  
\begin{eqnarray}
\label{asymptd2}
n_{+}^{(c)}(T)\lambda^2  \approx \left\{
\begin{array}{ccl}
\frac{\pi^2}{12\,\tilde{v}^2} & \textrm{for} & \tilde{v} \gg 1 \\
    &        &         \\
    -\,\log(4\tilde{v}^2) & \textrm{for} & \tilde{v} \ll 1 \;.
    \end{array}
    \right.
\end{eqnarray}  
Obviously, in contrast to $d>2$, for $d=2$ the dependence on $v$ does not drop out of the asymptotic formula for $n_+^{(c)}(T)$ in the regime $\tilde{v}\ll 1$.    
 
\subsection{Weyl-type coupling} 
In this section we briefly comment on another type of S-O coupling discussed in literature (see e.g.~\cite{Liao_2015, Wu_2017_2}) - the so-called Weyl-type coupling. 
The mean-field interacting Bose gas with Weyl S-O coupling is described by the Hamiltonian in Eq.~\eqref{Hamiltonian} with 
\begin{align}
\label{W01}
    \hat{\mathcal{H}}^{(1)}=\frac{\vec{p}^2}{2m}\hat{\mathbb{1}}_\sigma - v \,\vec{\hat{\sigma}} \cdot \vec{p} \,,
\end{align}
where $\hat{\vec{\sigma}}=(\hat{\sigma}_{1},\hat{\sigma}_{2},\hat{\sigma}_{3})$ and $\vec{p}=(p_1,p_2,\dots,p_d)$. After diagonalization analogous to the one described for the Rashba coupling one obtains the following single particle spectrum 
\begin{align}
\label{Wdisp01}
E^{(W)}_{\vec{p},\sigma}=\frac{\vec{p}^2}{2m}+\sigma v\sqrt{p_1^2+p_2^2+p_3^2}\;, 
\end{align} 
where $\sigma\in\{\pm\}$. Following the steps described previously for the Rashba coupling one obtains the following equations for the densities $n_{+}$ and $n_{-}$ corresponding to two branches of the dispersion relation: 
\begin{align} 
\label{WDensitieseq}
    n_+ \lambda^d &=g_{\frac{d}{2}}\left(e^{s_+}\right)- 2 \tilde{v} \int\limits_{v}^{\infty} dx \,g_{\frac{d-3}{2}}\left(e^{s_{+}+\tilde{v}^2-x^2}\right) 
    +\frac{1}{V}\frac{e^{s_+}}{1-e^{s_+}}\,, \nonumber\\  
    n_- \lambda^d &=g_{\frac{d}{2}}\left(e^{s_{-}-\tilde{v}^2}\right)+ 2 \sqrt{\pi}\tilde{v} g_{\frac{d-2}{2}} \left(e^{s_{-}}\right) - 2 \tilde{v} \int\limits_{v}^{\infty} dx\, g_{\frac{d-3}{2}}\left(e^{s_{-}-x^2}\right) \\ 
    &+\frac{\kappa}{V}\frac{e^{s_-}}{1-e^{s_-}}\;,     \nonumber
\end{align} 
where the variables $s_{\pm}$ are given by Eqs \eqref{s01} and \eqref{s02}. It follows from Eqs \eqref{WDensitieseq} that for  $\sigma=-1$ the maximum value of the thermal contribution to $n_{-}$ is infinite for $d\in\{3,4\}$ and finite for $d>4$. On the other hand, for $\sigma=+1$ the maximum value of the thermal contribution to $n_{+}$ remains finite for $d\in\{3,4\}$. This implies that in $d=3$ the component $\sigma=-1$ does not display the BE condensation while the component $\sigma=+1$ becomes BE condensed at low temperatures. We conclude that also for the Weyl-type S-O coupling, the modulated condensate does not represent a stable long-range ordered thermodynamic phase at any $T>0$ in the physical dimensionality $d=3$. 

In the following analysis of phase diagrams we shall restrict our discussion exclusively to the Rashba coupling. 
 \section{Phase diagrams} 
In the present section we focus on  $d=3$ and extract representative equilibrium phase diagrams by direct numerical evaluation of the saddle points of the function $\Phi(n_+, n_-)$ given by Eq.~(\ref{Phi}) and Eq.~(\ref{tton}). Whenever multiple solutions occur (as happens in certain cases analyzed below), the physically realized equilibrium corresponds to the one with the lowest value of $\Phi$. In our procedure, we treat the case $\tilde{v}=0$ considered in Ref.~\cite{Jakubczyk_2024} as a reference situation and  investigate how the phase diagram varies upon switching on the S-O coupling. In view of the results of the previous section, it is legitimate to expect a drastic change of the structure of the diagram even for tiny values of $\tilde{v}$. Note that the lower indices $1$ and $2$ used in \cite{Jakubczyk_2024} to denote the components of the binary mixture, their densities, chemical potentials and coupling constants are in the presently studied S-O coupled Bose gas replaced by the lower indices $+$ and $-$ corresponding to the two quasiparticle branches. 

We observe that occurrence of condensate of the quasiparticles from the upper branch implies, via Eq.~(\ref{Densitieseq}), that $s_+=0$, which yields a simple relation between $n_+$ and $n_-$ following from Eq.~(\ref{s01}). Substituting this into $\Phi (n_+, n_-)$ reduces the problem to saddle-point analysis of a function of a single variable, with a consistency requirement that $n_+$ exceeds the critical  value $n_+=n_+^{(c)}$. In contrast, in absence of condensation, the reduction stemming from Eq.~(\ref{s01}) at $s_{+}=0$ does not work, and one has to analyze  $\Phi(n_{+},n_{-})$ as a function of two variables. In the implemented procedure we consider both types of solutions $(n_{+},n_{-})$ corresponding to $n_{+} > n_{+}^{c}$ and to $n_{+} < n_{+}^{c}$. In certain cases studied  below (and illustrated in Fig.~4), in thermodynamic states close to the transition, we find competing solutions of both types and the dominant saddle point changes when varying the thermodynamic parameters ($\mu_+$, $\mu_-$, $\beta$), which results in a first-order condensation transition. 

Concerning the case $\tilde{v}=0$, it is well-known from literature  (see e.g.~\cite{Ao_1998, Ohberg_1998, Shi_2000, Riboli_2002, Altman_2003, Catani_2008, Tojo_2010, Facchi_2011, McCarron_2011, Schaeybroeck_2013, Ceccarelli_2015, Lingua_2015, Ceccarelli_2016, Utesov_2018, Boudjemaa_2018, Ota_2020, Spada_2023_2, Jakubczyk_2024}) that the structure of the phase diagram significantly depends on the sign of the interaction matrix determinant $D=a_+a_- -a_{\pm}^2$. For $D>0$, in addition to the normal phase,  there are three distinct thermodynamic phases hosting the Bose-Einstein condensates, hereafter referred to as BEC$_1$, BEC$_2$ and BEC$_{12}$. In the first two phases only one type of particles undergoes condensation, in the latter, both types. In contrast, for $D<0$ the system exhibits a first-order phase transition between the BEC$_1$ and BEC$_2$ phases and the BEC$_{12}$ phase does not occur in the phase diagram at all. The transition between the normal and BEC$_1$ or BEC$_2$ phases is always second order for $D>0$, but may be either first or second order \cite{Jakubczyk_2024} for $D<0$.  
For the present numerical analysis we chose the same sets of parameters, as investigated before in Ref.~\cite{Jakubczyk_2024}, which are close to those pertinent to ultracold gases of Lithium atoms. We chose an arbitrary, relatively small value of the S-O coupling ($\tilde{v}=1/10$).

In Fig.~3 we present our numerical results for a projection of the phase diagram on the $(\beta\mu_+, \beta\mu_-)$ plane in a situation, where $D>0$ and $\tilde{v}=1/10$. The system hosts only two phases (normal and BEC$_{+}$, representing the uniform condensate of upper-branch quasiparticles), separated by a second-order transition. The phase diagram is superimposed with the one pertinent to $\tilde{v}=0$ (thin dotted lines). As compared to the case $\tilde{v}=0$ the normal-BEC$_1$ transition line is uniformly shifted towards lower values of the chemical potential $\mu_+$ (corresponding also to lower values of the density $n_+$). The BEC$_2$ phase, which for $\tilde{v}>0$ would be represented by the modulated condensate BEC$_{-}$, is expelled from the phase diagram for $\tilde{v}>0$. The phase BEC$_{+-}$ is also absent.    
\begin{figure}
\includegraphics[width=8.5cm]{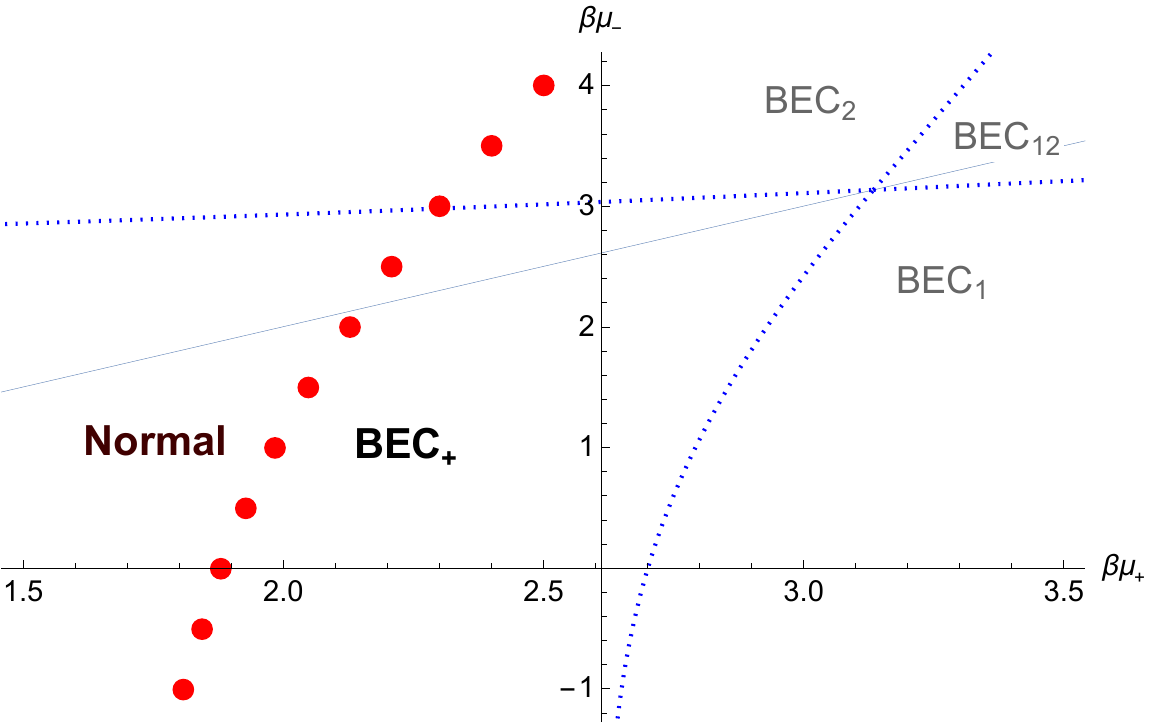}
\caption{A projection of the phase diagram on the ($\beta\mu_+$, $\beta\mu_-$) plane in a situation where $D>0$. The system hosts only two thermodynamic phases (normal and the uniform BEC$_+$ phase), separated by a second-order transition, which is represented by the red points. The plot parameters are as follows: $\beta a_+\lambda^{-3}= \beta a_-\lambda^{-3}=1$, $\beta a_{\pm}\lambda^{-3}=1/5$, $\tilde{v}=1/10$. The phase diagram is superimposed with the one corresponding to $\tilde{v}=0$, which displays the normal, BEC$_1$, BEC$_2$, and BEC$_{12}$ phases with the phase boundaries represented by the thin dotted lines. The auxiliary thin straight line corresponds to $\mu_+=\mu_-$. }
\end{figure}    

The situation exhibited in Fig.~3 is now contrasted with the one occurring for $D<0$, see Fig.~4 for a representative plot. Note that in the latter case and for $\tilde{v}=0$ condensation may occur either by a first- or a second-order transition.    
\begin{figure}
\includegraphics[width=8.5cm]{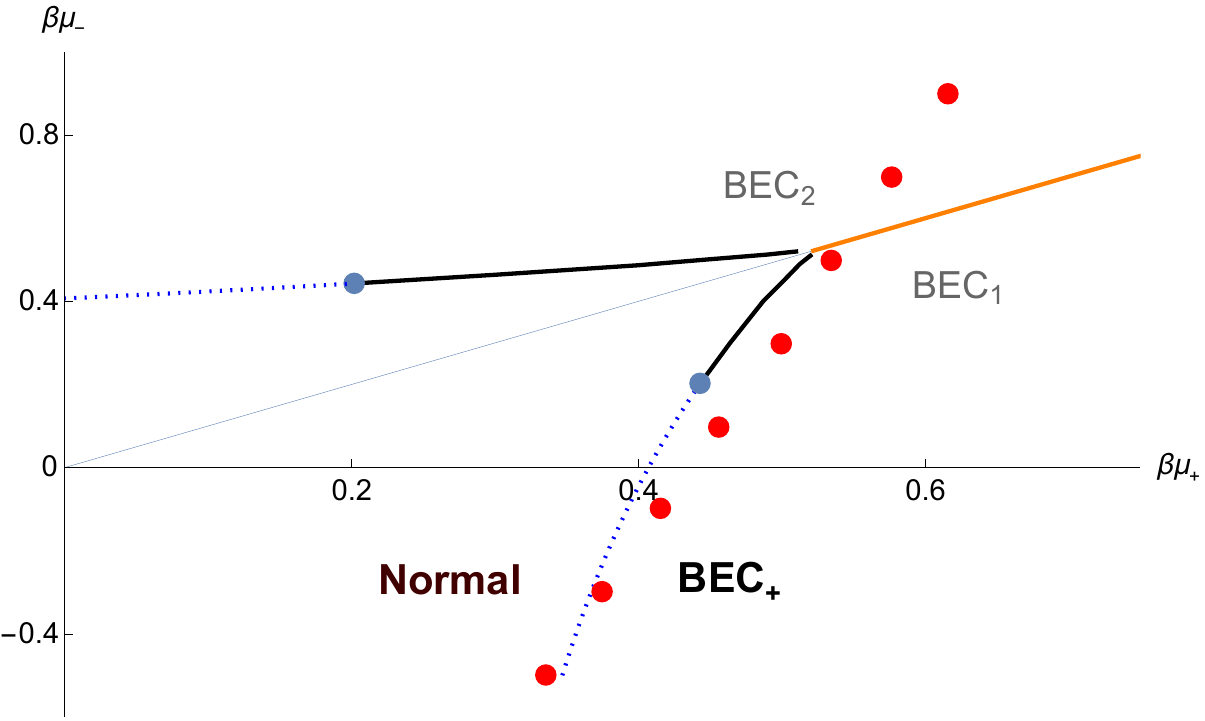}
\caption{A projection of the phase diagram on the ($\beta\mu_+$, $\beta\mu_-$) plane in a situation where $D<0$. The system hosts only two thermodynamic phases (normal and the uniform BEC$_+$ phase), separated by a first-order transition, which is represented by the red points. The plot parameters are as follows: $\beta a_+\lambda^{-3}= \beta a_-\lambda^{-3}=1/10$, $\beta a_{\pm}\lambda^{-3}=1/2$, $\tilde{v}=1/10$. The phase diagram is superimposed with the one corresponding to $\tilde{v}=0$, which displays the normal, BEC$_1$, and BEC$_2$ phases. The corresponding phase boundaries are represented by the thin  lines (dotted for second-order transitions and full for first-order transitions). The auxiliary thin straight line corresponds to $\mu_+=\mu_-$.} 
\end{figure}  
Even though in both the situations (represented by Fig.~3 and Fig.~4) the system displays (for $\tilde{v}>0$) the same two thermodynamic phases (normal and the uniform BEC$_+$ phase), 
the comparison reveals a rather striking difference between the impact of $\tilde{v}$ on the phase diagram. For the case illustrated in Fig.~4, the condensation transition is first-order and is accompanied with jumps of density of magnitude substantially larger than observed for $\tilde{v}=0$. In addition, the position of the phase boundary is not uniformly shifted and its qualitative shape is not retained upon switching on the coupling $\tilde{v}$. Instead, at least in the range of parameters we investigated, it follows a straight line located in the vicinity of the second-order phase boundary present for $\tilde{v}=0$. 

In summary, irrespective of the sign of $D$, for $T>0$ and $\tilde{v}>0$ we anticipate a phase diagram involving only the normal phase and the BEC$_+$ phase representing the uniform condensate of the upper-branch ($\sigma=+1$) quasiparticles. The structure of the phase diagram is therefore simpler than the one occurring for $\tilde{v}=0$. Within the  considered range of parameters we have numerically found that as long as $D>0$, the normal-BEC$_+$ transition retains its continuous character in presence of the S-O coupling (albeit it belongs to a different universality class as discussed in Sec.~II). For $D<0$ on the other hand, we have identified a tendency favouring and enhancing the first-order normal-BEC$_+$ transition.   

We finish this section with an empirical observation concerning the nature of the saddle-point of the function $\Phi(n_+,n_-)$ in the non-BEC phase, which turns out to be different depending on the sign of $D$. While for $D>0$ the saddle-point solution represents a maximum of $\Phi(n_+,n_-)$, it is a genuine saddle-point (neither maximum nor minimum) for $D<0$. The numerical procedure is significantly simpler for $v=0$ (see Ref.~\cite{Jakubczyk_2024}), where one can use one of the saddle-point equations~(\ref{Densitieseq}) to eliminate one of the densities, which is impossible for $v>0$.  
 
\section{Summary and outlook}
We have studied the effect of the presence of spin-orbit coupling $v$ on the $T>0$ phase diagram of the Bose gas  with interparticle interactions treated at mean-field level. The calculation of this paper exploits a connection to our recent studies of interacting Bose mixtures and allows in particular for resolving the limit $v\to 0$. Our analysis is exact within the proposed model and points at fragility of long-range ordered states characterized by ordering wavevector $\vec{Q}\neq \vec{0}$ in spatial dimensionality $d\leq 3$ at any temperature $T>0$ for both Rashba- and Weyl-type S-O couplings. On the other hand, presence of the S-O coupling  in dimensionality $d=3$ elevates the critical temperature for condensation into the uniform $\vec{Q}=\vec{0}$ state, while in $d=2$ it stabilizes the uniform long-range ordered condensate. The properties of the transition between the normal and uniform BEC phases are also strongly influenced by $v$ either by altering the universality class of the second-order transition, or by driving the transition first-order, as we demonstrated in Sec.~II and III. It is interesting to note, that while the modulated BEC condensates are stable only above dimensionality $d=4$ for Weyl S-O coupling, in the case of Rashba S-O coupling their instability occurs only for $d\leq 3$, such that the physical dimensionality $d=3$ coincides with the lower critical dimension. While we expect that the result concerning the instability of fully-developed modulated long-range order in $d=3$ should remain valid also beyond mean-field, we anticipate the possibility of realizing such phases in some form of algebraic order akin to the Kosterlitz-Thouless phase.  Stable long-ranged ordered condensates with $\vec{Q}\neq 0$ should on the other hand be achieved for anisotropic S-O couplings (as recognized in earlier literature).  

We finally point out, that the phase diagram at $T=0$ exhibits a much richer structure involving many possible long-ranged ordered phases separated by quantum phase transitions of different nature, such as, for example, the quantum Lifshitz points \cite{Gubbels_2009,  Wu_2017, Zdybel_2020}. At $T>0$ these are visible in the properties of different correlation functions, which constitutes an interesting and not much explored avenue for future research.

\begin{acknowledgments}
 We acknowledge support from the Polish National Science Center via grant2021/43/B/ST3/01223. P.J. thanks Oskar Stachowiak for useful discussions on related topics as well as for pointing out important references.
\end{acknowledgments}

\bibliography{bibliography.bib}

\begin{thebibliography}{74}%
\makeatletter
\providecommand \@ifxundefined [1]{%
 \@ifx{#1\undefined}
}%
\providecommand \@ifnum [1]{%
 \ifnum #1\expandafter \@firstoftwo
 \else \expandafter \@secondoftwo
 \fi
}%
\providecommand \@ifx [1]{%
 \ifx #1\expandafter \@firstoftwo
 \else \expandafter \@secondoftwo
 \fi
}%
\providecommand \natexlab [1]{#1}%
\providecommand \enquote  [1]{``#1''}%
\providecommand \bibnamefont  [1]{#1}%
\providecommand \bibfnamefont [1]{#1}%
\providecommand \citenamefont [1]{#1}%
\providecommand \href@noop [0]{\@secondoftwo}%
\providecommand \href [0]{\begingroup \@sanitize@url \@href}%
\providecommand \@href[1]{\@@startlink{#1}\@@href}%
\providecommand \@@href[1]{\endgroup#1\@@endlink}%
\providecommand \@sanitize@url [0]{\catcode `\\12\catcode `\$12\catcode
  `\&12\catcode `\#12\catcode `\^12\catcode `\_12\catcode `\%12\relax}%
\providecommand \@@startlink[1]{}%
\providecommand \@@endlink[0]{}%
\providecommand \url  [0]{\begingroup\@sanitize@url \@url }%
\providecommand \@url [1]{\endgroup\@href {#1}{\urlprefix }}%
\providecommand \urlprefix  [0]{URL }%
\providecommand \Eprint [0]{\href }%
\providecommand \doibase [0]{http://dx.doi.org/}%
\providecommand \selectlanguage [0]{\@gobble}%
\providecommand \bibinfo  [0]{\@secondoftwo}%
\providecommand \bibfield  [0]{\@secondoftwo}%
\providecommand \translation [1]{[#1]}%
\providecommand \BibitemOpen [0]{}%
\providecommand \bibitemStop [0]{}%
\providecommand \bibitemNoStop [0]{.\EOS\space}%
\providecommand \EOS [0]{\spacefactor3000\relax}%
\providecommand \BibitemShut  [1]{\csname bibitem#1\endcsname}%
\let\auto@bib@innerbib\@empty
\bibitem [{\citenamefont {Boninsegni}\ and\ \citenamefont
  {Prokof'ev}(2012)}]{Boninsegni_579}%
  \BibitemOpen
  \bibfield  {author} {\bibinfo {author} {\bibfnamefont {M.}~\bibnamefont
  {Boninsegni}}\ and\ \bibinfo {author} {\bibfnamefont {N.~V.}\ \bibnamefont
  {Prokof'ev}},\ }\href {\doibase 10.1103/RevModPhys.84.759} {\bibfield
  {journal} {\bibinfo  {journal} {Rev. Mod. Phys.}\ }\textbf {\bibinfo {volume}
  {84}},\ \bibinfo {pages} {759} (\bibinfo {year} {2012})}\BibitemShut
  {NoStop}%
\bibitem [{\citenamefont {Dalibard}\ \emph {et~al.}(2011)\citenamefont
  {Dalibard}, \citenamefont {Gerbier}, \citenamefont
  {Juzeli\ifmmode~\bar{u}\else \={u}\fi{}nas},\ and\ \citenamefont
  {\"Ohberg}}]{Dalibard_2011}%
  \BibitemOpen
  \bibfield  {author} {\bibinfo {author} {\bibfnamefont {J.}~\bibnamefont
  {Dalibard}}, \bibinfo {author} {\bibfnamefont {F.}~\bibnamefont {Gerbier}},
  \bibinfo {author} {\bibfnamefont {G.}~\bibnamefont
  {Juzeli\ifmmode~\bar{u}\else \={u}\fi{}nas}}, \ and\ \bibinfo {author}
  {\bibfnamefont {P.}~\bibnamefont {\"Ohberg}},\ }\href {\doibase
  10.1103/RevModPhys.83.1523} {\bibfield  {journal} {\bibinfo  {journal} {Rev.
  Mod. Phys.}\ }\textbf {\bibinfo {volume} {83}},\ \bibinfo {pages} {1523}
  (\bibinfo {year} {2011})}\BibitemShut {NoStop}%
\bibitem [{\citenamefont {Galitski}\ and\ \citenamefont
  {Spielman}(2013)}]{Galitski_2013}%
  \BibitemOpen
  \bibfield  {author} {\bibinfo {author} {\bibfnamefont {V.}~\bibnamefont
  {Galitski}}\ and\ \bibinfo {author} {\bibfnamefont {I.~B.}\ \bibnamefont
  {Spielman}},\ }\href@noop {} {\bibfield  {journal} {\bibinfo  {journal}
  {Nature}\ }\textbf {\bibinfo {volume} {494}},\ \bibinfo {pages} {49}
  (\bibinfo {year} {2013})}\BibitemShut {NoStop}%
\bibitem [{\citenamefont {Zhou}\ \emph {et~al.}(2013)\citenamefont {Zhou},
  \citenamefont {Li}, \citenamefont {Cai},\ and\ \citenamefont
  {Wu}}]{Zhou_2013}%
  \BibitemOpen
  \bibfield  {author} {\bibinfo {author} {\bibfnamefont {X.}~\bibnamefont
  {Zhou}}, \bibinfo {author} {\bibfnamefont {Y.}~\bibnamefont {Li}}, \bibinfo
  {author} {\bibfnamefont {Z.}~\bibnamefont {Cai}}, \ and\ \bibinfo {author}
  {\bibfnamefont {C.}~\bibnamefont {Wu}},\ }\href {\doibase
  10.1088/0953-4075/46/13/134001} {\bibfield  {journal} {\bibinfo  {journal}
  {Journal of Physics B: Atomic, Molecular and Optical Physics}\ }\textbf
  {\bibinfo {volume} {46}},\ \bibinfo {pages} {134001} (\bibinfo {year}
  {2013})}\BibitemShut {NoStop}%
\bibitem [{\citenamefont {Goldman}\ \emph {et~al.}(2014)\citenamefont
  {Goldman}, \citenamefont {Juzeliūnas}, \citenamefont {Öhberg},\ and\
  \citenamefont {Spielman}}]{Goldman_2014}%
  \BibitemOpen
  \bibfield  {author} {\bibinfo {author} {\bibfnamefont {N.}~\bibnamefont
  {Goldman}}, \bibinfo {author} {\bibfnamefont {G.}~\bibnamefont
  {Juzeliūnas}}, \bibinfo {author} {\bibfnamefont {P.}~\bibnamefont
  {Öhberg}}, \ and\ \bibinfo {author} {\bibfnamefont {I.~B.}\ \bibnamefont
  {Spielman}},\ }\href {\doibase 10.1088/0034-4885/77/12/126401} {\bibfield
  {journal} {\bibinfo  {journal} {Reports on Progress in Physics}\ }\textbf
  {\bibinfo {volume} {77}},\ \bibinfo {pages} {126401} (\bibinfo {year}
  {2014})}\BibitemShut {NoStop}%
\bibitem [{\citenamefont {Zhai}(2015)}]{Zhai_2015}%
  \BibitemOpen
  \bibfield  {author} {\bibinfo {author} {\bibfnamefont {H.}~\bibnamefont
  {Zhai}},\ }\href {\doibase 10.1088/0034-4885/78/2/026001} {\bibfield
  {journal} {\bibinfo  {journal} {Reports on Progress in Physics}\ }\textbf
  {\bibinfo {volume} {78}},\ \bibinfo {pages} {026001} (\bibinfo {year}
  {2015})}\BibitemShut {NoStop}%
\bibitem [{\citenamefont {Zhang}\ \emph {et~al.}(2016)\citenamefont {Zhang},
  \citenamefont {Mossman}, \citenamefont {Busch}, \citenamefont {Engels},\ and\
  \citenamefont {Zhang}}]{Zhang_2016}%
  \BibitemOpen
  \bibfield  {author} {\bibinfo {author} {\bibfnamefont {Y.}~\bibnamefont
  {Zhang}}, \bibinfo {author} {\bibfnamefont {M.~E.}\ \bibnamefont {Mossman}},
  \bibinfo {author} {\bibfnamefont {T.}~\bibnamefont {Busch}}, \bibinfo
  {author} {\bibfnamefont {P.}~\bibnamefont {Engels}}, \ and\ \bibinfo {author}
  {\bibfnamefont {C.}~\bibnamefont {Zhang}},\ }\href@noop {} {\bibfield
  {journal} {\bibinfo  {journal} {Frontiers of Physics}\ }\textbf {\bibinfo
  {volume} {11}},\ \bibinfo {pages} {118103} (\bibinfo {year}
  {2016})}\BibitemShut {NoStop}%
\bibitem [{\citenamefont {Recati}\ and\ \citenamefont
  {Stringari}(2022)}]{Recati_2022}%
  \BibitemOpen
  \bibfield  {author} {\bibinfo {author} {\bibfnamefont {A.}~\bibnamefont
  {Recati}}\ and\ \bibinfo {author} {\bibfnamefont {S.}~\bibnamefont
  {Stringari}},\ }\href {\doibase
  https://doi.org/10.1146/annurev-conmatphys-031820-121316} {\bibfield
  {journal} {\bibinfo  {journal} {Annual Review of Condensed Matter Physics}\
  }\textbf {\bibinfo {volume} {13}},\ \bibinfo {pages} {407} (\bibinfo {year}
  {2022})}\BibitemShut {NoStop}%
\bibitem [{\citenamefont {Stanescu}\ \emph {et~al.}(2008)\citenamefont
  {Stanescu}, \citenamefont {Anderson},\ and\ \citenamefont
  {Galitski}}]{Stanescu_2008}%
  \BibitemOpen
  \bibfield  {author} {\bibinfo {author} {\bibfnamefont {T.~D.}\ \bibnamefont
  {Stanescu}}, \bibinfo {author} {\bibfnamefont {B.}~\bibnamefont {Anderson}},
  \ and\ \bibinfo {author} {\bibfnamefont {V.}~\bibnamefont {Galitski}},\
  }\href {\doibase 10.1103/PhysRevA.78.023616} {\bibfield  {journal} {\bibinfo
  {journal} {Phys. Rev. A}\ }\textbf {\bibinfo {volume} {78}},\ \bibinfo
  {pages} {023616} (\bibinfo {year} {2008})}\BibitemShut {NoStop}%
\bibitem [{\citenamefont {Barnett}\ \emph
  {et~al.}(2012{\natexlab{a}})\citenamefont {Barnett}, \citenamefont {Powell},
  \citenamefont {Gra\ss{}}, \citenamefont {Lewenstein},\ and\ \citenamefont
  {Das~Sarma}}]{Barnett_2012}%
  \BibitemOpen
  \bibfield  {author} {\bibinfo {author} {\bibfnamefont {R.}~\bibnamefont
  {Barnett}}, \bibinfo {author} {\bibfnamefont {S.}~\bibnamefont {Powell}},
  \bibinfo {author} {\bibfnamefont {T.}~\bibnamefont {Gra\ss{}}}, \bibinfo
  {author} {\bibfnamefont {M.}~\bibnamefont {Lewenstein}}, \ and\ \bibinfo
  {author} {\bibfnamefont {S.}~\bibnamefont {Das~Sarma}},\ }\href {\doibase
  10.1103/PhysRevA.85.023615} {\bibfield  {journal} {\bibinfo  {journal} {Phys.
  Rev. A}\ }\textbf {\bibinfo {volume} {85}},\ \bibinfo {pages} {023615}
  (\bibinfo {year} {2012}{\natexlab{a}})}\BibitemShut {NoStop}%
\bibitem [{\citenamefont {Barnett}\ \emph
  {et~al.}(2012{\natexlab{b}})\citenamefont {Barnett}, \citenamefont {Powell},
  \citenamefont {Gra\ss{}}, \citenamefont {Lewenstein},\ and\ \citenamefont
  {Das~Sarma}}]{Barnett_2012_erratum}%
  \BibitemOpen
  \bibfield  {author} {\bibinfo {author} {\bibfnamefont {R.}~\bibnamefont
  {Barnett}}, \bibinfo {author} {\bibfnamefont {S.}~\bibnamefont {Powell}},
  \bibinfo {author} {\bibfnamefont {T.}~\bibnamefont {Gra\ss{}}}, \bibinfo
  {author} {\bibfnamefont {M.}~\bibnamefont {Lewenstein}}, \ and\ \bibinfo
  {author} {\bibfnamefont {S.}~\bibnamefont {Das~Sarma}},\ }\href {\doibase
  10.1103/PhysRevA.85.049905} {\bibfield  {journal} {\bibinfo  {journal} {Phys.
  Rev. A}\ }\textbf {\bibinfo {volume} {85}},\ \bibinfo {pages} {049905}
  (\bibinfo {year} {2012}{\natexlab{b}})}\BibitemShut {NoStop}%
\bibitem [{\citenamefont {Ozawa}\ and\ \citenamefont
  {Baym}(2012)}]{Ozawa_2012_2}%
  \BibitemOpen
  \bibfield  {author} {\bibinfo {author} {\bibfnamefont {T.}~\bibnamefont
  {Ozawa}}\ and\ \bibinfo {author} {\bibfnamefont {G.}~\bibnamefont {Baym}},\
  }\href {\doibase 10.1103/PhysRevLett.109.025301} {\bibfield  {journal}
  {\bibinfo  {journal} {Phys. Rev. Lett.}\ }\textbf {\bibinfo {volume} {109}},\
  \bibinfo {pages} {025301} (\bibinfo {year} {2012})}\BibitemShut {NoStop}%
\bibitem [{\citenamefont {Cui}\ and\ \citenamefont {Zhou}(2013)}]{Cui_2013}%
  \BibitemOpen
  \bibfield  {author} {\bibinfo {author} {\bibfnamefont {X.}~\bibnamefont
  {Cui}}\ and\ \bibinfo {author} {\bibfnamefont {Q.}~\bibnamefont {Zhou}},\
  }\href {\doibase 10.1103/PhysRevA.87.031604} {\bibfield  {journal} {\bibinfo
  {journal} {Phys. Rev. A}\ }\textbf {\bibinfo {volume} {87}},\ \bibinfo
  {pages} {031604} (\bibinfo {year} {2013})}\BibitemShut {NoStop}%
\bibitem [{\citenamefont {Yu}(2013)}]{Yu_2013}%
  \BibitemOpen
  \bibfield  {author} {\bibinfo {author} {\bibfnamefont {Z.-Q.}\ \bibnamefont
  {Yu}},\ }\href {\doibase 10.1103/PhysRevA.87.051606} {\bibfield  {journal}
  {\bibinfo  {journal} {Phys. Rev. A}\ }\textbf {\bibinfo {volume} {87}},\
  \bibinfo {pages} {051606} (\bibinfo {year} {2013})}\BibitemShut {NoStop}%
\bibitem [{\citenamefont {He}\ \emph {et~al.}(2013)\citenamefont {He},
  \citenamefont {You},\ and\ \citenamefont {Liu}}]{He_2013}%
  \BibitemOpen
  \bibfield  {author} {\bibinfo {author} {\bibfnamefont {P.-S.}\ \bibnamefont
  {He}}, \bibinfo {author} {\bibfnamefont {W.-L.}\ \bibnamefont {You}}, \ and\
  \bibinfo {author} {\bibfnamefont {W.-M.}\ \bibnamefont {Liu}},\ }\href
  {\doibase 10.1103/PhysRevA.87.063603} {\bibfield  {journal} {\bibinfo
  {journal} {Phys. Rev. A}\ }\textbf {\bibinfo {volume} {87}},\ \bibinfo
  {pages} {063603} (\bibinfo {year} {2013})}\BibitemShut {NoStop}%
\bibitem [{\citenamefont {Liao}\ \emph {et~al.}(2014)\citenamefont {Liao},
  \citenamefont {Huang}, \citenamefont {Lin},\ and\ \citenamefont
  {Fialko}}]{Liao_2014}%
  \BibitemOpen
  \bibfield  {author} {\bibinfo {author} {\bibfnamefont {R.}~\bibnamefont
  {Liao}}, \bibinfo {author} {\bibfnamefont {Z.-G.}\ \bibnamefont {Huang}},
  \bibinfo {author} {\bibfnamefont {X.-M.}\ \bibnamefont {Lin}}, \ and\
  \bibinfo {author} {\bibfnamefont {O.}~\bibnamefont {Fialko}},\ }\href
  {\doibase 10.1103/PhysRevA.89.063614} {\bibfield  {journal} {\bibinfo
  {journal} {Phys. Rev. A}\ }\textbf {\bibinfo {volume} {89}},\ \bibinfo
  {pages} {063614} (\bibinfo {year} {2014})}\BibitemShut {NoStop}%
\bibitem [{\citenamefont {Roy}\ \emph {et~al.}(2014)\citenamefont {Roy},
  \citenamefont {Ray},\ and\ \citenamefont {Sinha}}]{Roy_2014}%
  \BibitemOpen
  \bibfield  {author} {\bibinfo {author} {\bibfnamefont {A.}~\bibnamefont
  {Roy}}, \bibinfo {author} {\bibfnamefont {S.}~\bibnamefont {Ray}}, \ and\
  \bibinfo {author} {\bibfnamefont {S.}~\bibnamefont {Sinha}},\ }\href
  {\doibase 10.1140/epjd/e2014-50394-9} {\bibfield  {journal} {\bibinfo
  {journal} {The European Physical Journal D}\ }\textbf {\bibinfo {volume}
  {68}},\ \bibinfo {pages} {376} (\bibinfo {year} {2014})}\BibitemShut
  {NoStop}%
\bibitem [{\citenamefont {Hickey}\ and\ \citenamefont
  {Paramekanti}(2014)}]{Hickey_2014}%
  \BibitemOpen
  \bibfield  {author} {\bibinfo {author} {\bibfnamefont {C.}~\bibnamefont
  {Hickey}}\ and\ \bibinfo {author} {\bibfnamefont {A.}~\bibnamefont
  {Paramekanti}},\ }\href {\doibase 10.1103/PhysRevLett.113.265302} {\bibfield
  {journal} {\bibinfo  {journal} {Phys. Rev. Lett.}\ }\textbf {\bibinfo
  {volume} {113}},\ \bibinfo {pages} {265302} (\bibinfo {year}
  {2014})}\BibitemShut {NoStop}%
\bibitem [{\citenamefont {Mardonov}\ \emph {et~al.}(2015)\citenamefont
  {Mardonov}, \citenamefont {Sherman}, \citenamefont {Muga}, \citenamefont
  {Wang}, \citenamefont {Ban},\ and\ \citenamefont {Chen}}]{Mardonov_2015}%
  \BibitemOpen
  \bibfield  {author} {\bibinfo {author} {\bibfnamefont {S.}~\bibnamefont
  {Mardonov}}, \bibinfo {author} {\bibfnamefont {E.~Y.}\ \bibnamefont
  {Sherman}}, \bibinfo {author} {\bibfnamefont {J.~G.}\ \bibnamefont {Muga}},
  \bibinfo {author} {\bibfnamefont {H.-W.}\ \bibnamefont {Wang}}, \bibinfo
  {author} {\bibfnamefont {Y.}~\bibnamefont {Ban}}, \ and\ \bibinfo {author}
  {\bibfnamefont {X.}~\bibnamefont {Chen}},\ }\href {\doibase
  10.1103/PhysRevA.91.043604} {\bibfield  {journal} {\bibinfo  {journal} {Phys.
  Rev. A}\ }\textbf {\bibinfo {volume} {91}},\ \bibinfo {pages} {043604}
  (\bibinfo {year} {2015})}\BibitemShut {NoStop}%
\bibitem [{\citenamefont {Liao}\ \emph {et~al.}(2015)\citenamefont {Liao},
  \citenamefont {Fialko}, \citenamefont {Brand},\ and\ \citenamefont
  {Z\"ulicke}}]{Liao_2015}%
  \BibitemOpen
  \bibfield  {author} {\bibinfo {author} {\bibfnamefont {R.}~\bibnamefont
  {Liao}}, \bibinfo {author} {\bibfnamefont {O.}~\bibnamefont {Fialko}},
  \bibinfo {author} {\bibfnamefont {J.}~\bibnamefont {Brand}}, \ and\ \bibinfo
  {author} {\bibfnamefont {U.}~\bibnamefont {Z\"ulicke}},\ }\href {\doibase
  10.1103/PhysRevA.92.043633} {\bibfield  {journal} {\bibinfo  {journal} {Phys.
  Rev. A}\ }\textbf {\bibinfo {volume} {92}},\ \bibinfo {pages} {043633}
  (\bibinfo {year} {2015})}\BibitemShut {NoStop}%
\bibitem [{\citenamefont {Galteland}\ and\ \citenamefont
  {Sudb\o{}}(2016)}]{Galteland_2016}%
  \BibitemOpen
  \bibfield  {author} {\bibinfo {author} {\bibfnamefont {P.~N.}\ \bibnamefont
  {Galteland}}\ and\ \bibinfo {author} {\bibfnamefont {A.}~\bibnamefont
  {Sudb\o{}}},\ }\href {\doibase 10.1103/PhysRevB.94.054510} {\bibfield
  {journal} {\bibinfo  {journal} {Phys. Rev. B}\ }\textbf {\bibinfo {volume}
  {94}},\ \bibinfo {pages} {054510} (\bibinfo {year} {2016})}\BibitemShut
  {NoStop}%
\bibitem [{\citenamefont {Wu}\ \emph {et~al.}(2017)\citenamefont {Wu},
  \citenamefont {Zhou},\ and\ \citenamefont {Wu}}]{Wu_2017}%
  \BibitemOpen
  \bibfield  {author} {\bibinfo {author} {\bibfnamefont {J.}~\bibnamefont
  {Wu}}, \bibinfo {author} {\bibfnamefont {F.}~\bibnamefont {Zhou}}, \ and\
  \bibinfo {author} {\bibfnamefont {C.}~\bibnamefont {Wu}},\ }\href {\doibase
  10.1103/PhysRevB.96.085140} {\bibfield  {journal} {\bibinfo  {journal} {Phys.
  Rev. B}\ }\textbf {\bibinfo {volume} {96}},\ \bibinfo {pages} {085140}
  (\bibinfo {year} {2017})}\BibitemShut {NoStop}%
\bibitem [{\citenamefont {Kawasaki}\ and\ \citenamefont
  {Holzmann}(2017)}]{Kawasaki_2017}%
  \BibitemOpen
  \bibfield  {author} {\bibinfo {author} {\bibfnamefont {E.}~\bibnamefont
  {Kawasaki}}\ and\ \bibinfo {author} {\bibfnamefont {M.}~\bibnamefont
  {Holzmann}},\ }\href {\doibase 10.1103/PhysRevA.95.051601} {\bibfield
  {journal} {\bibinfo  {journal} {Phys. Rev. A}\ }\textbf {\bibinfo {volume}
  {95}},\ \bibinfo {pages} {051601} (\bibinfo {year} {2017})}\BibitemShut
  {NoStop}%
\bibitem [{\citenamefont {Yang}(2021)}]{Yang_2021}%
  \BibitemOpen
  \bibfield  {author} {\bibinfo {author} {\bibfnamefont {L.}~\bibnamefont
  {Yang}},\ }\href {\doibase 10.1103/PhysRevA.104.023320} {\bibfield  {journal}
  {\bibinfo  {journal} {Phys. Rev. A}\ }\textbf {\bibinfo {volume} {104}},\
  \bibinfo {pages} {023320} (\bibinfo {year} {2021})}\BibitemShut {NoStop}%
\bibitem [{\citenamefont {Shimahara}(1998)}]{Shimahara_1998}%
  \BibitemOpen
  \bibfield  {author} {\bibinfo {author} {\bibfnamefont {H.}~\bibnamefont
  {Shimahara}},\ }\href@noop {} {\bibfield  {journal} {\bibinfo  {journal}
  {Journal of the Physical Society of Japan}\ }\textbf {\bibinfo {volume}
  {67}},\ \bibinfo {pages} {1872} (\bibinfo {year} {1998})}\BibitemShut
  {NoStop}%
\bibitem [{\citenamefont {Radzihovsky}\ and\ \citenamefont
  {Vishwanath}(2009)}]{Radzihovshy_2009}%
  \BibitemOpen
  \bibfield  {author} {\bibinfo {author} {\bibfnamefont {L.}~\bibnamefont
  {Radzihovsky}}\ and\ \bibinfo {author} {\bibfnamefont {A.}~\bibnamefont
  {Vishwanath}},\ }\href {\doibase 10.1103/PhysRevLett.103.010404} {\bibfield
  {journal} {\bibinfo  {journal} {Phys. Rev. Lett.}\ }\textbf {\bibinfo
  {volume} {103}},\ \bibinfo {pages} {010404} (\bibinfo {year}
  {2009})}\BibitemShut {NoStop}%
\bibitem [{\citenamefont {Radzihovsky}(2011)}]{Radzihovsky_2011}%
  \BibitemOpen
  \bibfield  {author} {\bibinfo {author} {\bibfnamefont {L.}~\bibnamefont
  {Radzihovsky}},\ }\href {\doibase 10.1103/PhysRevA.84.023611} {\bibfield
  {journal} {\bibinfo  {journal} {Phys. Rev. A}\ }\textbf {\bibinfo {volume}
  {84}},\ \bibinfo {pages} {023611} (\bibinfo {year} {2011})}\BibitemShut
  {NoStop}%
\bibitem [{\citenamefont {Yin}\ \emph {et~al.}(2014)\citenamefont {Yin},
  \citenamefont {Martikainen},\ and\ \citenamefont {T\"orm\"a}}]{Yin_2014}%
  \BibitemOpen
  \bibfield  {author} {\bibinfo {author} {\bibfnamefont {S.}~\bibnamefont
  {Yin}}, \bibinfo {author} {\bibfnamefont {J.-P.}\ \bibnamefont
  {Martikainen}}, \ and\ \bibinfo {author} {\bibfnamefont {P.}~\bibnamefont
  {T\"orm\"a}},\ }\href {\doibase 10.1103/PhysRevB.89.014507} {\bibfield
  {journal} {\bibinfo  {journal} {Phys. Rev. B}\ }\textbf {\bibinfo {volume}
  {89}},\ \bibinfo {pages} {014507} (\bibinfo {year} {2014})}\BibitemShut
  {NoStop}%
\bibitem [{\citenamefont {Jakubczyk}(2017)}]{Jakubczyk_2017}%
  \BibitemOpen
  \bibfield  {author} {\bibinfo {author} {\bibfnamefont {P.}~\bibnamefont
  {Jakubczyk}},\ }\href {\doibase 10.1103/PhysRevA.95.063626} {\bibfield
  {journal} {\bibinfo  {journal} {Phys. Rev. A}\ }\textbf {\bibinfo {volume}
  {95}},\ \bibinfo {pages} {063626} (\bibinfo {year} {2017})}\BibitemShut
  {NoStop}%
\bibitem [{\citenamefont {Wang}\ \emph {et~al.}(2018)\citenamefont {Wang},
  \citenamefont {Che}, \citenamefont {Zhang},\ and\ \citenamefont
  {Chen}}]{Wang_2018}%
  \BibitemOpen
  \bibfield  {author} {\bibinfo {author} {\bibfnamefont {J.}~\bibnamefont
  {Wang}}, \bibinfo {author} {\bibfnamefont {Y.}~\bibnamefont {Che}}, \bibinfo
  {author} {\bibfnamefont {L.}~\bibnamefont {Zhang}}, \ and\ \bibinfo {author}
  {\bibfnamefont {Q.}~\bibnamefont {Chen}},\ }\href {\doibase
  10.1103/PhysRevB.97.134513} {\bibfield  {journal} {\bibinfo  {journal} {Phys.
  Rev. B}\ }\textbf {\bibinfo {volume} {97}},\ \bibinfo {pages} {134513}
  (\bibinfo {year} {2018})}\BibitemShut {NoStop}%
\bibitem [{\citenamefont {Zdybel}\ \emph {et~al.}(2021)\citenamefont {Zdybel},
  \citenamefont {Homenda}, \citenamefont {Chlebicki},\ and\ \citenamefont
  {Jakubczyk}}]{Zdybel_2021}%
  \BibitemOpen
  \bibfield  {author} {\bibinfo {author} {\bibfnamefont {P.}~\bibnamefont
  {Zdybel}}, \bibinfo {author} {\bibfnamefont {M.}~\bibnamefont {Homenda}},
  \bibinfo {author} {\bibfnamefont {A.}~\bibnamefont {Chlebicki}}, \ and\
  \bibinfo {author} {\bibfnamefont {P.}~\bibnamefont {Jakubczyk}},\ }\href
  {\doibase 10.1103/PhysRevA.104.063317} {\bibfield  {journal} {\bibinfo
  {journal} {Phys. Rev. A}\ }\textbf {\bibinfo {volume} {104}},\ \bibinfo
  {pages} {063317} (\bibinfo {year} {2021})}\BibitemShut {NoStop}%
\bibitem [{\citenamefont {Pini}\ \emph {et~al.}(2021)\citenamefont {Pini},
  \citenamefont {Pieri},\ and\ \citenamefont {Calvanese~Strinati}}]{Pini_2021}%
  \BibitemOpen
  \bibfield  {author} {\bibinfo {author} {\bibfnamefont {M.}~\bibnamefont
  {Pini}}, \bibinfo {author} {\bibfnamefont {P.}~\bibnamefont {Pieri}}, \ and\
  \bibinfo {author} {\bibfnamefont {G.}~\bibnamefont {Calvanese~Strinati}},\
  }\href {\doibase 10.1103/PhysRevResearch.3.043068} {\bibfield  {journal}
  {\bibinfo  {journal} {Phys. Rev. Res.}\ }\textbf {\bibinfo {volume} {3}},\
  \bibinfo {pages} {043068} (\bibinfo {year} {2021})}\BibitemShut {NoStop}%
\bibitem [{\citenamefont {Pini}\ \emph {et~al.}(2023)\citenamefont {Pini},
  \citenamefont {Pieri},\ and\ \citenamefont {Calvanese~Strinati}}]{Pini_2023}%
  \BibitemOpen
  \bibfield  {author} {\bibinfo {author} {\bibfnamefont {M.}~\bibnamefont
  {Pini}}, \bibinfo {author} {\bibfnamefont {P.}~\bibnamefont {Pieri}}, \ and\
  \bibinfo {author} {\bibfnamefont {G.}~\bibnamefont {Calvanese~Strinati}},\
  }\href {\doibase 10.1103/PhysRevB.107.054505} {\bibfield  {journal} {\bibinfo
   {journal} {Phys. Rev. B}\ }\textbf {\bibinfo {volume} {107}},\ \bibinfo
  {pages} {054505} (\bibinfo {year} {2023})}\BibitemShut {NoStop}%
\bibitem [{\citenamefont {Osterloh}\ \emph {et~al.}(2005)\citenamefont
  {Osterloh}, \citenamefont {Baig}, \citenamefont {Santos}, \citenamefont
  {Zoller},\ and\ \citenamefont {Lewenstein}}]{Osterloh_2005}%
  \BibitemOpen
  \bibfield  {author} {\bibinfo {author} {\bibfnamefont {K.}~\bibnamefont
  {Osterloh}}, \bibinfo {author} {\bibfnamefont {M.}~\bibnamefont {Baig}},
  \bibinfo {author} {\bibfnamefont {L.}~\bibnamefont {Santos}}, \bibinfo
  {author} {\bibfnamefont {P.}~\bibnamefont {Zoller}}, \ and\ \bibinfo {author}
  {\bibfnamefont {M.}~\bibnamefont {Lewenstein}},\ }\href {\doibase
  10.1103/PhysRevLett.95.010403} {\bibfield  {journal} {\bibinfo  {journal}
  {Phys. Rev. Lett.}\ }\textbf {\bibinfo {volume} {95}},\ \bibinfo {pages}
  {010403} (\bibinfo {year} {2005})}\BibitemShut {NoStop}%
\bibitem [{\citenamefont {Zhu}\ \emph {et~al.}(2006)\citenamefont {Zhu},
  \citenamefont {Fu}, \citenamefont {Wu}, \citenamefont {Zhang},\ and\
  \citenamefont {Duan}}]{Zhu_2006}%
  \BibitemOpen
  \bibfield  {author} {\bibinfo {author} {\bibfnamefont {S.-L.}\ \bibnamefont
  {Zhu}}, \bibinfo {author} {\bibfnamefont {H.}~\bibnamefont {Fu}}, \bibinfo
  {author} {\bibfnamefont {C.-J.}\ \bibnamefont {Wu}}, \bibinfo {author}
  {\bibfnamefont {S.-C.}\ \bibnamefont {Zhang}}, \ and\ \bibinfo {author}
  {\bibfnamefont {L.-M.}\ \bibnamefont {Duan}},\ }\href {\doibase
  10.1103/PhysRevLett.97.240401} {\bibfield  {journal} {\bibinfo  {journal}
  {Phys. Rev. Lett.}\ }\textbf {\bibinfo {volume} {97}},\ \bibinfo {pages}
  {240401} (\bibinfo {year} {2006})}\BibitemShut {NoStop}%
\bibitem [{\citenamefont {Satija}\ \emph {et~al.}(2006)\citenamefont {Satija},
  \citenamefont {Dakin},\ and\ \citenamefont {Clark}}]{Satija_2006}%
  \BibitemOpen
  \bibfield  {author} {\bibinfo {author} {\bibfnamefont {I.~I.}\ \bibnamefont
  {Satija}}, \bibinfo {author} {\bibfnamefont {D.~C.}\ \bibnamefont {Dakin}}, \
  and\ \bibinfo {author} {\bibfnamefont {C.~W.}\ \bibnamefont {Clark}},\ }\href
  {\doibase 10.1103/PhysRevLett.97.216401} {\bibfield  {journal} {\bibinfo
  {journal} {Phys. Rev. Lett.}\ }\textbf {\bibinfo {volume} {97}},\ \bibinfo
  {pages} {216401} (\bibinfo {year} {2006})}\BibitemShut {NoStop}%
\bibitem [{\citenamefont {Stanescu}\ \emph {et~al.}(2007)\citenamefont
  {Stanescu}, \citenamefont {Zhang},\ and\ \citenamefont
  {Galitski}}]{Stanescu_2007}%
  \BibitemOpen
  \bibfield  {author} {\bibinfo {author} {\bibfnamefont {T.~D.}\ \bibnamefont
  {Stanescu}}, \bibinfo {author} {\bibfnamefont {C.}~\bibnamefont {Zhang}}, \
  and\ \bibinfo {author} {\bibfnamefont {V.}~\bibnamefont {Galitski}},\ }\href
  {\doibase 10.1103/PhysRevLett.99.110403} {\bibfield  {journal} {\bibinfo
  {journal} {Phys. Rev. Lett.}\ }\textbf {\bibinfo {volume} {99}},\ \bibinfo
  {pages} {110403} (\bibinfo {year} {2007})}\BibitemShut {NoStop}%
\bibitem [{\citenamefont {Hemmer}\ and\ \citenamefont
  {Lebowitz}(1976)}]{Hemmer_1976}%
  \BibitemOpen
  \bibfield  {author} {\bibinfo {author} {\bibfnamefont {P.~C.}\ \bibnamefont
  {Hemmer}}\ and\ \bibinfo {author} {\bibfnamefont {J.~L.}\ \bibnamefont
  {Lebowitz}},\ }in\ \href@noop {} {\emph {\bibinfo {booktitle} {Phase
  Transitions and Critical Phenomena, vol. 5b}}},\ \bibinfo {editor} {edited
  by\ \bibinfo {editor} {\bibfnamefont {C.}~\bibnamefont {Domb}}\ and\ \bibinfo
  {editor} {\bibfnamefont {M.~S.}\ \bibnamefont {Green}}}\ (\bibinfo
  {publisher} {Academic Press},\ \bibinfo {address} {London, New York, San
  Francisco},\ \bibinfo {year} {1976})\ pp.\ \bibinfo {pages}
  {107--203}\BibitemShut {NoStop}%
\bibitem [{\citenamefont {Davies}(1972)}]{Davies_1972}%
  \BibitemOpen
  \bibfield  {author} {\bibinfo {author} {\bibfnamefont {E.~B.}\ \bibnamefont
  {Davies}},\ }\href@noop {} {\bibfield  {journal} {\bibinfo  {journal}
  {Communications in Mathematical Physics}\ }\textbf {\bibinfo {volume} {28}},\
  \bibinfo {pages} {69 } (\bibinfo {year} {1972})}\BibitemShut {NoStop}%
\bibitem [{\citenamefont {Buffet}\ and\ \citenamefont
  {Pulè}(1983)}]{Buffet_1983}%
  \BibitemOpen
  \bibfield  {author} {\bibinfo {author} {\bibfnamefont {E.}~\bibnamefont
  {Buffet}}\ and\ \bibinfo {author} {\bibfnamefont {J.~V.}\ \bibnamefont
  {Pulè}},\ }\href {\doibase 10.1063/1.525855} {\bibfield  {journal} {\bibinfo
   {journal} {Journal of Mathematical Physics}\ }\textbf {\bibinfo {volume}
  {24}},\ \bibinfo {pages} {1608} (\bibinfo {year} {1983})}\BibitemShut
  {NoStop}%
\bibitem [{\citenamefont {van~den Berg}\ \emph {et~al.}(1984)\citenamefont
  {van~den Berg}, \citenamefont {Lewis},\ and\ \citenamefont
  {de~Smedt}}]{Berg_1984}%
  \BibitemOpen
  \bibfield  {author} {\bibinfo {author} {\bibfnamefont {M.}~\bibnamefont
  {van~den Berg}}, \bibinfo {author} {\bibfnamefont {J.~T.}\ \bibnamefont
  {Lewis}}, \ and\ \bibinfo {author} {\bibfnamefont {P.}~\bibnamefont
  {de~Smedt}},\ }\href {\doibase 10.1007/BF01010502} {\bibfield  {journal}
  {\bibinfo  {journal} {Journal of Statistical Physics}\ }\textbf {\bibinfo
  {volume} {37}},\ \bibinfo {pages} {697} (\bibinfo {year} {1984})}\BibitemShut
  {NoStop}%
\bibitem [{\citenamefont {Lewis}(1985)}]{Lewis_1985}%
  \BibitemOpen
  \bibfield  {author} {\bibinfo {author} {\bibfnamefont {J.~T.}\ \bibnamefont
  {Lewis}},\ }in\ \href@noop {} {\emph {\bibinfo {booktitle} {Statistical
  Mechanics and Field Theory: Mathematical Aspects}}},\ \bibinfo {editor}
  {edited by\ \bibinfo {editor} {\bibfnamefont {T.~C.}\ \bibnamefont {Dorlas}},
  \bibinfo {editor} {\bibfnamefont {N.~M.}\ \bibnamefont {Hugenholtz}}, \ and\
  \bibinfo {editor} {\bibfnamefont {M.}~\bibnamefont {Winnink}}}\ (\bibinfo
  {publisher} {Springer Berlin Heidelberg},\ \bibinfo {address} {Berlin,
  Heidelberg},\ \bibinfo {year} {1985})\ pp.\ \bibinfo {pages}
  {234--256}\BibitemShut {NoStop}%
\bibitem [{\citenamefont {de~Smedt}(1986)}]{Smedt_1986}%
  \BibitemOpen
  \bibfield  {author} {\bibinfo {author} {\bibfnamefont {P.}~\bibnamefont
  {de~Smedt}},\ }\href {\doibase 10.1007/BF01033087} {\bibfield  {journal}
  {\bibinfo  {journal} {Journal of Statistical Physics}\ }\textbf {\bibinfo
  {volume} {45}},\ \bibinfo {pages} {201} (\bibinfo {year} {1986})}\BibitemShut
  {NoStop}%
\bibitem [{\citenamefont {Zagrebnov}\ and\ \citenamefont
  {Bru}(2001)}]{Zagrebnov_2001}%
  \BibitemOpen
  \bibfield  {author} {\bibinfo {author} {\bibfnamefont {V.~A.}\ \bibnamefont
  {Zagrebnov}}\ and\ \bibinfo {author} {\bibfnamefont {J.-B.}\ \bibnamefont
  {Bru}},\ }\href {\doibase https://doi.org/10.1016/S0370-1573(00)00132-0}
  {\bibfield  {journal} {\bibinfo  {journal} {Physics Reports}\ }\textbf
  {\bibinfo {volume} {350}},\ \bibinfo {pages} {291} (\bibinfo {year}
  {2001})}\BibitemShut {NoStop}%
\bibitem [{\citenamefont {Napi\'orkowski}\ and\ \citenamefont
  {Piasecki}(2011)}]{Napiorkowski_2011}%
  \BibitemOpen
  \bibfield  {author} {\bibinfo {author} {\bibfnamefont {M.}~\bibnamefont
  {Napi\'orkowski}}\ and\ \bibinfo {author} {\bibfnamefont {J.}~\bibnamefont
  {Piasecki}},\ }\href {\doibase 10.1103/PhysRevE.84.061105} {\bibfield
  {journal} {\bibinfo  {journal} {Phys. Rev. E}\ }\textbf {\bibinfo {volume}
  {84}},\ \bibinfo {pages} {061105} (\bibinfo {year} {2011})}\BibitemShut
  {NoStop}%
\bibitem [{\citenamefont {Napiórkowski}\ \emph {et~al.}(2013)\citenamefont
  {Napiórkowski}, \citenamefont {Jakubczyk},\ and\ \citenamefont
  {Nowak}}]{Napiorkowski_2013}%
  \BibitemOpen
  \bibfield  {author} {\bibinfo {author} {\bibfnamefont {M.}~\bibnamefont
  {Napiórkowski}}, \bibinfo {author} {\bibfnamefont {P.}~\bibnamefont
  {Jakubczyk}}, \ and\ \bibinfo {author} {\bibfnamefont {K.}~\bibnamefont
  {Nowak}},\ }\href {\doibase 10.1088/1742-5468/2013/06/P06015} {\bibfield
  {journal} {\bibinfo  {journal} {Journal of Statistical Mechanics: Theory and
  Experiment}\ }\textbf {\bibinfo {volume} {2013}},\ \bibinfo {pages} {P06015}
  (\bibinfo {year} {2013})}\BibitemShut {NoStop}%
\bibitem [{\citenamefont {Diehl}\ and\ \citenamefont
  {Rutkevich}(2017)}]{Diehl_2017}%
  \BibitemOpen
  \bibfield  {author} {\bibinfo {author} {\bibfnamefont {H.~W.}\ \bibnamefont
  {Diehl}}\ and\ \bibinfo {author} {\bibfnamefont {S.~B.}\ \bibnamefont
  {Rutkevich}},\ }\href {\doibase 10.1103/PhysRevE.95.062112} {\bibfield
  {journal} {\bibinfo  {journal} {Phys. Rev. E}\ }\textbf {\bibinfo {volume}
  {95}},\ \bibinfo {pages} {062112} (\bibinfo {year} {2017})}\BibitemShut
  {NoStop}%
\bibitem [{\citenamefont {Jakubczyk}\ and\ \citenamefont
  {Wojtkiewicz}(2018)}]{Jakubczyk_2018}%
  \BibitemOpen
  \bibfield  {author} {\bibinfo {author} {\bibfnamefont {P.}~\bibnamefont
  {Jakubczyk}}\ and\ \bibinfo {author} {\bibfnamefont {J.}~\bibnamefont
  {Wojtkiewicz}},\ }\href {\doibase 10.1088/1742-5468/aabc7c} {\bibfield
  {journal} {\bibinfo  {journal} {Journal of Statistical Mechanics: Theory and
  Experiment}\ }\textbf {\bibinfo {volume} {2018}},\ \bibinfo {pages} {053105}
  (\bibinfo {year} {2018})}\BibitemShut {NoStop}%
\bibitem [{\citenamefont {Berlin}\ and\ \citenamefont
  {Kac}(1952)}]{Berlin_1952}%
  \BibitemOpen
  \bibfield  {author} {\bibinfo {author} {\bibfnamefont {T.~H.}\ \bibnamefont
  {Berlin}}\ and\ \bibinfo {author} {\bibfnamefont {M.}~\bibnamefont {Kac}},\
  }\href {\doibase 10.1103/PhysRev.86.821} {\bibfield  {journal} {\bibinfo
  {journal} {Phys. Rev.}\ }\textbf {\bibinfo {volume} {86}},\ \bibinfo {pages}
  {821} (\bibinfo {year} {1952})}\BibitemShut {NoStop}%
\bibitem [{\citenamefont {Stanley}(1968)}]{Stanley_1968}%
  \BibitemOpen
  \bibfield  {author} {\bibinfo {author} {\bibfnamefont {H.~E.}\ \bibnamefont
  {Stanley}},\ }\href {\doibase 10.1103/PhysRev.176.718} {\bibfield  {journal}
  {\bibinfo  {journal} {Phys. Rev.}\ }\textbf {\bibinfo {volume} {176}},\
  \bibinfo {pages} {718} (\bibinfo {year} {1968})}\BibitemShut {NoStop}%
\bibitem [{\citenamefont {Moshe}\ and\ \citenamefont
  {Zinn-Justin}(2003)}]{Moshe_2003}%
  \BibitemOpen
  \bibfield  {author} {\bibinfo {author} {\bibfnamefont {M.}~\bibnamefont
  {Moshe}}\ and\ \bibinfo {author} {\bibfnamefont {J.}~\bibnamefont
  {Zinn-Justin}},\ }\href {\doibase
  https://doi.org/10.1016/S0370-1573(03)00263-1} {\bibfield  {journal}
  {\bibinfo  {journal} {Physics Reports}\ }\textbf {\bibinfo {volume} {385}},\
  \bibinfo {pages} {69} (\bibinfo {year} {2003})}\BibitemShut {NoStop}%
\bibitem [{\citenamefont {Jakubczyk}\ \emph {et~al.}(2024)\citenamefont
  {Jakubczyk}, \citenamefont {My\ifmmode~\acute{s}\else \'{s}\fi{}liwy},\ and\
  \citenamefont {Napi\'orkowski}}]{Jakubczyk_2024}%
  \BibitemOpen
  \bibfield  {author} {\bibinfo {author} {\bibfnamefont {P.}~\bibnamefont
  {Jakubczyk}}, \bibinfo {author} {\bibfnamefont {K.}~\bibnamefont
  {My\ifmmode~\acute{s}\else \'{s}\fi{}liwy}}, \ and\ \bibinfo {author}
  {\bibfnamefont {M.}~\bibnamefont {Napi\'orkowski}},\ }\href {\doibase
  10.1103/PhysRevA.109.013312} {\bibfield  {journal} {\bibinfo  {journal}
  {Phys. Rev. A}\ }\textbf {\bibinfo {volume} {109}},\ \bibinfo {pages}
  {013312} (\bibinfo {year} {2024})}\BibitemShut {NoStop}%
\bibitem [{\citenamefont {Ziff}\ \emph {et~al.}(1977)\citenamefont {Ziff},
  \citenamefont {Uhlenbeck},\ and\ \citenamefont {Kac}}]{Ziff_1977}%
  \BibitemOpen
  \bibfield  {author} {\bibinfo {author} {\bibfnamefont {R.~M.}\ \bibnamefont
  {Ziff}}, \bibinfo {author} {\bibfnamefont {G.~E.}\ \bibnamefont {Uhlenbeck}},
  \ and\ \bibinfo {author} {\bibfnamefont {M.}~\bibnamefont {Kac}},\ }\href
  {\doibase https://doi.org/10.1016/0370-1573(77)90052-7} {\bibfield  {journal}
  {\bibinfo  {journal} {Physics Reports}\ }\textbf {\bibinfo {volume} {32}},\
  \bibinfo {pages} {169} (\bibinfo {year} {1977})}\BibitemShut {NoStop}%
\bibitem [{\citenamefont {Kardar}(2007)}]{Kardar_book}%
  \BibitemOpen
  \bibfield  {author} {\bibinfo {author} {\bibfnamefont {M.}~\bibnamefont
  {Kardar}},\ }\href {https://books.google.pl/books?id=nTxBhGX01P4C} {\emph
  {\bibinfo {title} {Statistical Physics of Particles}}}\ (\bibinfo
  {publisher} {Cambridge University Press},\ \bibinfo {year}
  {2007})\BibitemShut {NoStop}%
\bibitem [{\citenamefont {Wu}\ and\ \citenamefont {Liao}(2017)}]{Wu_2017_2}%
  \BibitemOpen
  \bibfield  {author} {\bibinfo {author} {\bibfnamefont {T.}~\bibnamefont
  {Wu}}\ and\ \bibinfo {author} {\bibfnamefont {R.}~\bibnamefont {Liao}},\
  }\href {\doibase 10.1088/1367-2630/aa559b} {\bibfield  {journal} {\bibinfo
  {journal} {New Journal of Physics}\ }\textbf {\bibinfo {volume} {19}},\
  \bibinfo {pages} {013008} (\bibinfo {year} {2017})}\BibitemShut {NoStop}%
\bibitem [{\citenamefont {Ao}\ and\ \citenamefont {Chui}(1998)}]{Ao_1998}%
  \BibitemOpen
  \bibfield  {author} {\bibinfo {author} {\bibfnamefont {P.}~\bibnamefont
  {Ao}}\ and\ \bibinfo {author} {\bibfnamefont {S.~T.}\ \bibnamefont {Chui}},\
  }\href {\doibase 10.1103/PhysRevA.58.4836} {\bibfield  {journal} {\bibinfo
  {journal} {Phys. Rev. A}\ }\textbf {\bibinfo {volume} {58}},\ \bibinfo
  {pages} {4836} (\bibinfo {year} {1998})}\BibitemShut {NoStop}%
\bibitem [{\citenamefont {\"Ohberg}\ and\ \citenamefont
  {Stenholm}(1998)}]{Ohberg_1998}%
  \BibitemOpen
  \bibfield  {author} {\bibinfo {author} {\bibfnamefont {P.}~\bibnamefont
  {\"Ohberg}}\ and\ \bibinfo {author} {\bibfnamefont {S.}~\bibnamefont
  {Stenholm}},\ }\href {\doibase 10.1103/PhysRevA.57.1272} {\bibfield
  {journal} {\bibinfo  {journal} {Phys. Rev. A}\ }\textbf {\bibinfo {volume}
  {57}},\ \bibinfo {pages} {1272} (\bibinfo {year} {1998})}\BibitemShut
  {NoStop}%
\bibitem [{\citenamefont {Shi}\ \emph {et~al.}(2000)\citenamefont {Shi},
  \citenamefont {Zheng},\ and\ \citenamefont {Chui}}]{Shi_2000}%
  \BibitemOpen
  \bibfield  {author} {\bibinfo {author} {\bibfnamefont {H.}~\bibnamefont
  {Shi}}, \bibinfo {author} {\bibfnamefont {W.-M.}\ \bibnamefont {Zheng}}, \
  and\ \bibinfo {author} {\bibfnamefont {S.-T.}\ \bibnamefont {Chui}},\ }\href
  {\doibase 10.1103/PhysRevA.61.063613} {\bibfield  {journal} {\bibinfo
  {journal} {Phys. Rev. A}\ }\textbf {\bibinfo {volume} {61}},\ \bibinfo
  {pages} {063613} (\bibinfo {year} {2000})}\BibitemShut {NoStop}%
\bibitem [{\citenamefont {Riboli}\ and\ \citenamefont
  {Modugno}(2002)}]{Riboli_2002}%
  \BibitemOpen
  \bibfield  {author} {\bibinfo {author} {\bibfnamefont {F.}~\bibnamefont
  {Riboli}}\ and\ \bibinfo {author} {\bibfnamefont {M.}~\bibnamefont
  {Modugno}},\ }\href {\doibase 10.1103/PhysRevA.65.063614} {\bibfield
  {journal} {\bibinfo  {journal} {Phys. Rev. A}\ }\textbf {\bibinfo {volume}
  {65}},\ \bibinfo {pages} {063614} (\bibinfo {year} {2002})}\BibitemShut
  {NoStop}%
\bibitem [{\citenamefont {Altman}\ \emph {et~al.}(2003)\citenamefont {Altman},
  \citenamefont {Hofstetter}, \citenamefont {Demler},\ and\ \citenamefont
  {Lukin}}]{Altman_2003}%
  \BibitemOpen
  \bibfield  {author} {\bibinfo {author} {\bibfnamefont {E.}~\bibnamefont
  {Altman}}, \bibinfo {author} {\bibfnamefont {W.}~\bibnamefont {Hofstetter}},
  \bibinfo {author} {\bibfnamefont {E.}~\bibnamefont {Demler}}, \ and\ \bibinfo
  {author} {\bibfnamefont {M.~D.}\ \bibnamefont {Lukin}},\ }\href {\doibase
  10.1088/1367-2630/5/1/113} {\bibfield  {journal} {\bibinfo  {journal} {New
  Journal of Physics}\ }\textbf {\bibinfo {volume} {5}},\ \bibinfo {pages}
  {113} (\bibinfo {year} {2003})}\BibitemShut {NoStop}%
\bibitem [{\citenamefont {Catani}\ \emph {et~al.}(2008)\citenamefont {Catani},
  \citenamefont {De~Sarlo}, \citenamefont {Barontini}, \citenamefont
  {Minardi},\ and\ \citenamefont {Inguscio}}]{Catani_2008}%
  \BibitemOpen
  \bibfield  {author} {\bibinfo {author} {\bibfnamefont {J.}~\bibnamefont
  {Catani}}, \bibinfo {author} {\bibfnamefont {L.}~\bibnamefont {De~Sarlo}},
  \bibinfo {author} {\bibfnamefont {G.}~\bibnamefont {Barontini}}, \bibinfo
  {author} {\bibfnamefont {F.}~\bibnamefont {Minardi}}, \ and\ \bibinfo
  {author} {\bibfnamefont {M.}~\bibnamefont {Inguscio}},\ }\href {\doibase
  10.1103/PhysRevA.77.011603} {\bibfield  {journal} {\bibinfo  {journal} {Phys.
  Rev. A}\ }\textbf {\bibinfo {volume} {77}},\ \bibinfo {pages} {011603}
  (\bibinfo {year} {2008})}\BibitemShut {NoStop}%
\bibitem [{\citenamefont {Tojo}\ \emph {et~al.}(2010)\citenamefont {Tojo},
  \citenamefont {Taguchi}, \citenamefont {Masuyama}, \citenamefont {Hayashi},
  \citenamefont {Saito},\ and\ \citenamefont {Hirano}}]{Tojo_2010}%
  \BibitemOpen
  \bibfield  {author} {\bibinfo {author} {\bibfnamefont {S.}~\bibnamefont
  {Tojo}}, \bibinfo {author} {\bibfnamefont {Y.}~\bibnamefont {Taguchi}},
  \bibinfo {author} {\bibfnamefont {Y.}~\bibnamefont {Masuyama}}, \bibinfo
  {author} {\bibfnamefont {T.}~\bibnamefont {Hayashi}}, \bibinfo {author}
  {\bibfnamefont {H.}~\bibnamefont {Saito}}, \ and\ \bibinfo {author}
  {\bibfnamefont {T.}~\bibnamefont {Hirano}},\ }\href {\doibase
  10.1103/PhysRevA.82.033609} {\bibfield  {journal} {\bibinfo  {journal} {Phys.
  Rev. A}\ }\textbf {\bibinfo {volume} {82}},\ \bibinfo {pages} {033609}
  (\bibinfo {year} {2010})}\BibitemShut {NoStop}%
\bibitem [{\citenamefont {Facchi}\ \emph {et~al.}(2011)\citenamefont {Facchi},
  \citenamefont {Florio}, \citenamefont {Pascazio},\ and\ \citenamefont
  {Pepe}}]{Facchi_2011}%
  \BibitemOpen
  \bibfield  {author} {\bibinfo {author} {\bibfnamefont {P.}~\bibnamefont
  {Facchi}}, \bibinfo {author} {\bibfnamefont {G.}~\bibnamefont {Florio}},
  \bibinfo {author} {\bibfnamefont {S.}~\bibnamefont {Pascazio}}, \ and\
  \bibinfo {author} {\bibfnamefont {F.~V.}\ \bibnamefont {Pepe}},\ }\href
  {\doibase 10.1088/1751-8113/44/50/505305} {\bibfield  {journal} {\bibinfo
  {journal} {Journal of Physics A: Mathematical and Theoretical}\ }\textbf
  {\bibinfo {volume} {44}},\ \bibinfo {pages} {505305} (\bibinfo {year}
  {2011})}\BibitemShut {NoStop}%
\bibitem [{\citenamefont {McCarron}\ \emph {et~al.}(2011)\citenamefont
  {McCarron}, \citenamefont {Cho}, \citenamefont {Jenkin}, \citenamefont
  {K\"oppinger},\ and\ \citenamefont {Cornish}}]{McCarron_2011}%
  \BibitemOpen
  \bibfield  {author} {\bibinfo {author} {\bibfnamefont {D.~J.}\ \bibnamefont
  {McCarron}}, \bibinfo {author} {\bibfnamefont {H.~W.}\ \bibnamefont {Cho}},
  \bibinfo {author} {\bibfnamefont {D.~L.}\ \bibnamefont {Jenkin}}, \bibinfo
  {author} {\bibfnamefont {M.~P.}\ \bibnamefont {K\"oppinger}}, \ and\ \bibinfo
  {author} {\bibfnamefont {S.~L.}\ \bibnamefont {Cornish}},\ }\href {\doibase
  10.1103/PhysRevA.84.011603} {\bibfield  {journal} {\bibinfo  {journal} {Phys.
  Rev. A}\ }\textbf {\bibinfo {volume} {84}},\ \bibinfo {pages} {011603}
  (\bibinfo {year} {2011})}\BibitemShut {NoStop}%
\bibitem [{\citenamefont {{Van Schaeybroeck}}(2013)}]{Schaeybroeck_2013}%
  \BibitemOpen
  \bibfield  {author} {\bibinfo {author} {\bibfnamefont {B.}~\bibnamefont {{Van
  Schaeybroeck}}},\ }\href {\doibase
  https://doi.org/10.1016/j.physa.2013.04.026} {\bibfield  {journal} {\bibinfo
  {journal} {Physica A: Statistical Mechanics and its Applications}\ }\textbf
  {\bibinfo {volume} {392}},\ \bibinfo {pages} {3806} (\bibinfo {year}
  {2013})}\BibitemShut {NoStop}%
\bibitem [{\citenamefont {Ceccarelli}\ \emph {et~al.}(2015)\citenamefont
  {Ceccarelli}, \citenamefont {Nespolo}, \citenamefont {Pelissetto},\ and\
  \citenamefont {Vicari}}]{Ceccarelli_2015}%
  \BibitemOpen
  \bibfield  {author} {\bibinfo {author} {\bibfnamefont {G.}~\bibnamefont
  {Ceccarelli}}, \bibinfo {author} {\bibfnamefont {J.}~\bibnamefont {Nespolo}},
  \bibinfo {author} {\bibfnamefont {A.}~\bibnamefont {Pelissetto}}, \ and\
  \bibinfo {author} {\bibfnamefont {E.}~\bibnamefont {Vicari}},\ }\href
  {\doibase 10.1103/PhysRevA.92.043613} {\bibfield  {journal} {\bibinfo
  {journal} {Phys. Rev. A}\ }\textbf {\bibinfo {volume} {92}},\ \bibinfo
  {pages} {043613} (\bibinfo {year} {2015})}\BibitemShut {NoStop}%
\bibitem [{\citenamefont {Lingua}\ \emph {et~al.}(2015)\citenamefont {Lingua},
  \citenamefont {Guglielmino}, \citenamefont {Penna},\ and\ \citenamefont
  {Capogrosso~Sansone}}]{Lingua_2015}%
  \BibitemOpen
  \bibfield  {author} {\bibinfo {author} {\bibfnamefont {F.}~\bibnamefont
  {Lingua}}, \bibinfo {author} {\bibfnamefont {M.}~\bibnamefont {Guglielmino}},
  \bibinfo {author} {\bibfnamefont {V.}~\bibnamefont {Penna}}, \ and\ \bibinfo
  {author} {\bibfnamefont {B.}~\bibnamefont {Capogrosso~Sansone}},\ }\href
  {\doibase 10.1103/PhysRevA.92.053610} {\bibfield  {journal} {\bibinfo
  {journal} {Phys. Rev. A}\ }\textbf {\bibinfo {volume} {92}},\ \bibinfo
  {pages} {053610} (\bibinfo {year} {2015})}\BibitemShut {NoStop}%
\bibitem [{\citenamefont {Ceccarelli}\ \emph {et~al.}(2016)\citenamefont
  {Ceccarelli}, \citenamefont {Nespolo}, \citenamefont {Pelissetto},\ and\
  \citenamefont {Vicari}}]{Ceccarelli_2016}%
  \BibitemOpen
  \bibfield  {author} {\bibinfo {author} {\bibfnamefont {G.}~\bibnamefont
  {Ceccarelli}}, \bibinfo {author} {\bibfnamefont {J.}~\bibnamefont {Nespolo}},
  \bibinfo {author} {\bibfnamefont {A.}~\bibnamefont {Pelissetto}}, \ and\
  \bibinfo {author} {\bibfnamefont {E.}~\bibnamefont {Vicari}},\ }\href
  {\doibase 10.1103/PhysRevA.93.033647} {\bibfield  {journal} {\bibinfo
  {journal} {Phys. Rev. A}\ }\textbf {\bibinfo {volume} {93}},\ \bibinfo
  {pages} {033647} (\bibinfo {year} {2016})}\BibitemShut {NoStop}%
\bibitem [{\citenamefont {Utesov}\ \emph {et~al.}(2018)\citenamefont {Utesov},
  \citenamefont {Baglay},\ and\ \citenamefont {Andreev}}]{Utesov_2018}%
  \BibitemOpen
  \bibfield  {author} {\bibinfo {author} {\bibfnamefont {O.~I.}\ \bibnamefont
  {Utesov}}, \bibinfo {author} {\bibfnamefont {M.~I.}\ \bibnamefont {Baglay}},
  \ and\ \bibinfo {author} {\bibfnamefont {S.~V.}\ \bibnamefont {Andreev}},\
  }\href {\doibase 10.1103/PhysRevA.97.053617} {\bibfield  {journal} {\bibinfo
  {journal} {Phys. Rev. A}\ }\textbf {\bibinfo {volume} {97}},\ \bibinfo
  {pages} {053617} (\bibinfo {year} {2018})}\BibitemShut {NoStop}%
\bibitem [{\citenamefont {Boudjem\^aa}(2018)}]{Boudjemaa_2018}%
  \BibitemOpen
  \bibfield  {author} {\bibinfo {author} {\bibfnamefont {A.}~\bibnamefont
  {Boudjem\^aa}},\ }\href {\doibase 10.1103/PhysRevA.97.033627} {\bibfield
  {journal} {\bibinfo  {journal} {Phys. Rev. A}\ }\textbf {\bibinfo {volume}
  {97}},\ \bibinfo {pages} {033627} (\bibinfo {year} {2018})}\BibitemShut
  {NoStop}%
\bibitem [{\citenamefont {Ota}\ and\ \citenamefont
  {Giorgini}(2020)}]{Ota_2020}%
  \BibitemOpen
  \bibfield  {author} {\bibinfo {author} {\bibfnamefont {M.}~\bibnamefont
  {Ota}}\ and\ \bibinfo {author} {\bibfnamefont {S.}~\bibnamefont {Giorgini}},\
  }\href {\doibase 10.1103/PhysRevA.102.063303} {\bibfield  {journal} {\bibinfo
   {journal} {Phys. Rev. A}\ }\textbf {\bibinfo {volume} {102}},\ \bibinfo
  {pages} {063303} (\bibinfo {year} {2020})}\BibitemShut {NoStop}%
\bibitem [{\citenamefont {Spada}\ \emph {et~al.}(2023)\citenamefont {Spada},
  \citenamefont {Parisi}, \citenamefont {Pascual}, \citenamefont {Parker},
  \citenamefont {Billam}, \citenamefont {Pilati}, \citenamefont {Boronat},\
  and\ \citenamefont {Giorgini}}]{Spada_2023_2}%
  \BibitemOpen
  \bibfield  {author} {\bibinfo {author} {\bibfnamefont {G.}~\bibnamefont
  {Spada}}, \bibinfo {author} {\bibfnamefont {L.}~\bibnamefont {Parisi}},
  \bibinfo {author} {\bibfnamefont {G.}~\bibnamefont {Pascual}}, \bibinfo
  {author} {\bibfnamefont {N.~G.}\ \bibnamefont {Parker}}, \bibinfo {author}
  {\bibfnamefont {T.~P.}\ \bibnamefont {Billam}}, \bibinfo {author}
  {\bibfnamefont {S.}~\bibnamefont {Pilati}}, \bibinfo {author} {\bibfnamefont
  {J.}~\bibnamefont {Boronat}}, \ and\ \bibinfo {author} {\bibfnamefont
  {S.}~\bibnamefont {Giorgini}},\ }\href {\doibase
  10.21468/SciPostPhys.15.4.171} {\bibfield  {journal} {\bibinfo  {journal}
  {SciPost Phys.}\ }\textbf {\bibinfo {volume} {15}},\ \bibinfo {pages} {171}
  (\bibinfo {year} {2023})}\BibitemShut {NoStop}%
\bibitem [{\citenamefont {Gubbels}\ \emph {et~al.}(2009)\citenamefont
  {Gubbels}, \citenamefont {Baarsma},\ and\ \citenamefont
  {Stoof}}]{Gubbels_2009}%
  \BibitemOpen
  \bibfield  {author} {\bibinfo {author} {\bibfnamefont {K.~B.}\ \bibnamefont
  {Gubbels}}, \bibinfo {author} {\bibfnamefont {J.~E.}\ \bibnamefont
  {Baarsma}}, \ and\ \bibinfo {author} {\bibfnamefont {H.~T.~C.}\ \bibnamefont
  {Stoof}},\ }\href {\doibase 10.1103/PhysRevLett.103.195301} {\bibfield
  {journal} {\bibinfo  {journal} {Phys. Rev. Lett.}\ }\textbf {\bibinfo
  {volume} {103}},\ \bibinfo {pages} {195301} (\bibinfo {year}
  {2009})}\BibitemShut {NoStop}%
\bibitem [{\citenamefont {Zdybel}\ and\ \citenamefont
  {Jakubczyk}(2020)}]{Zdybel_2020}%
  \BibitemOpen
  \bibfield  {author} {\bibinfo {author} {\bibfnamefont {P.}~\bibnamefont
  {Zdybel}}\ and\ \bibinfo {author} {\bibfnamefont {P.}~\bibnamefont
  {Jakubczyk}},\ }\href {\doibase 10.1103/PhysRevResearch.2.033486} {\bibfield
  {journal} {\bibinfo  {journal} {Phys. Rev. Res.}\ }\textbf {\bibinfo {volume}
  {2}},\ \bibinfo {pages} {033486} (\bibinfo {year} {2020})}\BibitemShut
  {NoStop}%
\end{thebibliography}%
\bibliographystyle{apsrev4-1}

\end{document}